\documentclass[12pt]{article}
\usepackage{graphics,graphicx}
\usepackage{amssymb,epsfig,amsmath,euscript,array}
\usepackage{cite}
\usepackage{axodraw}
\usepackage{pstricks}
\usepackage{color}

\makeatletter
\@addtoreset{equation}{section}
\makeatother

\newcounter{multieqs}

\newcommand{\be}{\begin{equation}}
\newcommand{\ee}{\end{equation}}

\newcommand{\bm}[1]{\mbox{\boldmath $#1$}}

\newcommand{\kslash}{k \!\!\! / }

\newcommand{\lslash}{l \!\! / }
\newcommand{\Pslash}{P \!\!\!\! / }

\newcommand{\islash}{i \!\!\! / }
\newcommand{\jslash}{j \!\!\! / }
\newcommand{\aslash}{a \!\!\! / }
\newcommand{\bslash}{{b \hspace{-6pt} \slash} }

\newcommand{\onslash}{1 \!\!\! / }
\newcommand{\twslash}{2 \!\!\!/ }
\newcommand{\thslash}{3 \!\!\!/ }
\newcommand{\foslash}{4 \!\!\! / }
\newcommand{\fislash}{5 \!\!\! / }

\newcommand{\mslash}{m \!\!\! / }

\def\bd{\begin{document}}
\def\ed{\end{document}}
\def\nn{\nonumber}
\def\bea{\begin{eqnarray}}
\def\eea{\end{eqnarray}}

\def\ab{(ijab)}
\def\ba{(ijba)}
\def\ijab{{\tr}_{+}(\islash\, \jslash\, \aslash \, \bslash)}
\def\ijba{{\tr}_{+}(\islash\, \jslash\, \bslash \, \aslash)}
\def\ijaP{{\tr}_{+}(\islash\, \jslash\, \aslash \, \Pslash)}
\def\ijPLa{{\tr}_{+}(\islash\, \jslash\, \Pslash_L \, \aslash)}
\def\ijaPL{{\tr}_{+}(\islash\, \jslash\, \aslash \, \Pslash_L)}
\def\ijPLza{{\tr}_{+}(\islash\, \jslash\, \Pslash_{L;z} \, \aslash)}
\def\ijaPLz{{\tr}_{+}(\islash\, \jslash\, \aslash \, \Pslash_{L;z})}
\def\ijPa{{\tr}_{+}(\islash\, \jslash\, \Pslash \, \aslash)}
\def\iaPb{{\tr}_{+}(\islash\, \aslash\, \Pslash \, \bslash)}
\def\ibPa{{\tr}_{+}(\islash\, \bslash\, \Pslash \, \aslash)}
\def\ijPmu{{\tr}_{+}(\islash\, \jslash\, \Pslash \, \mu)}
\def\ibmuP{{\tr}_{+}(\islash\, \bslash\, \mu \, \Pslash)}
\def\ibmua{{\tr}_{+}(\islash\, \bslash\, \mu \, \aslash)}
\def\iamub{{\tr}_{+}(\islash\, \aslash\, \mu \, \bslash)}
\def\jaPb{{\tr}_{+}(\jslash\, \aslash\, \Pslash \, \bslash)}
\def\ijmuP{{\tr}_{+}(\islash\, \jslash\, \mu \, \Pslash)}
\def\ijmum{{\tr}_{+}(\islash\, \jslash\, \mu \, \mslash)}
\def\ijmmu{{\tr}_{+}(\islash\, \jslash\, \mslash \, \mu)}
\def\ijmP{{\tr}_{+}(\islash\, \jslash\, \mslash \, \Pslash)}
\def\iabP{{\tr}_{+}(\islash\, \aslash\, \bslash \, \Pslash)}
\def\ijbP{{\tr}_{+}(\islash\, \jslash\, \bslash \, \Pslash)}
\def\jbPa{{\tr}_{+}(\jslash\, \bslash\, \Pslash \, \aslash)}
\def\ijPb{{\tr}_{+}(\islash\, \jslash\, \Pslash \, \bslash)}
\def\jbmua{{\tr}_{+}(\jslash\, \bslash\, \mu \, \aslash)}

\def\loablt{ {\tr}_{+}(\lslash_1\, \aslash \, \bslash\, \lslash_2)}

\def\ijlolt{{\tr}_{+}(\islash\, \jslash\, \lslash_1 \, \lslash_2)}
\def\ijltlo{{\tr}_{+}(\islash\, \jslash\, \lslash_2 \, \lslash_1)}
\def\ibloa{{\tr}_{+}(\islash\, \bslash\, \lslash_1 \, \aslash)}
\def\jaltb{{\tr}_{+}(\jslash\, \aslash\, \lslash_2 \, \bslash)}
\def\ialtb{{\tr}_{+}(\islash\, \aslash\, \lslash_2 \, \bslash)}
\def\bltloa{{\tr}_{+}(\bslash\, \lslash_2\, \lslash_1 \, \aslash)}
\def\jbloa{{\tr}_{+}(\jslash\, \bslash\, \lslash_1 \, \aslash)}
\def\ibPb{{\tr}_{+}(\islash\, \bslash\, \Pslash \, \bslash)}
\def\ijltb{{\tr}_{+}(\islash\, \jslash\, \lslash_2 \, \bslash)}

\def\ijloa{{\tr}_{+}(\islash\, \jslash\,  \lslash_1 \, \aslash)}
\def\ijblt{{\tr}_{+}(\islash\, \jslash\,  \bslash \, \lslash_2)}

\def\jakb{{\tr}_{+}(\jslash\, \aslash\, \kslash \, \bslash)}
\def\iakb{{\tr}_{+}(\islash\, \aslash\, \kslash \, \bslash)}

\def\tofo{{\tr}_{+}(\onslash\, \thslash\, \twslash \, \foslash)}
\def\foto{{\tr}_{+}(\onslash\, \thslash\, \foslash \, \twslash)}
\def\tofi{{\tr}_{+}(\onslash\, \thslash\, \twslash \, \fislash)}
\def\fito{{\tr}_{+}(\onslash\, \thslash\, \fislash \, \twslash)}

\def\lrangle#1#2{\langle #1\,#2\rangle}

\def\Li{{$\rm Li}_2$}
\def\eps{\epsilon}
\def\epsuv{{\epsilon_{\rm \mbox{\tiny UV}}}}
\let\bm=\bibitem
\let\la=\label

\def\npb#1#2#3{Nucl. Phys. {\bf{B#1}} #3 (#2)}
\def\plb#1#2#3{Phys. Lett. {\bf{#1B}} #3 (#2)}
\def\prl#1#2#3{Phys. Rev. Lett. {\bf{#1}} #3 (#2)}
\def\prd#1#2#3{Phys. Rev. {D \bf{#1}} #3 (#2)}
\def\cmp#1#2#3{Comm. Math. Phys. {\bf{#1}} #3 (#2)}
\def\cqg#1#2#3{Class. Quantum Grav. {\bf{#1}} #3 (#2)}
\def\nppsa#1#2#3{Nucl. Phys. B (Proc. Suppl.) {\bf{#1A}}#3 (#2)}
\def\ap#1#2#3{Ann. of Phys. {\bf{#1}} #3 (#2)}
\def\ijmp#1#2#3{Int. J. Mod. Phys. {\bf{A#1}} #3 (#2)}
\def\rmp#1#2#3{Rev. Mod. Phys. {\bf{#1}} #3 (#2)}
\def\mpla#1#2#3{Mod. Phys. Lett. {\bf A#1} #3 (#2)}
\def\jhep#1#2#3{J. High Energy Phys. {\bf #1} #3 (#2)}
\def\atmp#1#2#3{Adv. Theor. Math. Phys. {\bf #1} #3 (#2)}
\newcommand{\EQ}[1]{\begin{equation} #1 \end{equation}}
\newcommand{\AL}[1]{\begin{subequations}\begin{align} #1 \end{align}\end{subequations}}
\newcommand{\SP}[1]{\begin{equation}\begin{split} #1 \end{split}\end{equation}}
\newcommand{\ALAT}[2]{\begin{subequations}\begin{alignat}{#1} #2 \end{alignat}
                        \end{subequations}}
\def\beqa{\begin{eqnarray}}
\def\eeqa{\end{eqnarray}}
\def\beq{\begin{equation}}
\def\eeq{\end{equation}}
\def\sst{\scriptscriptstyle}
\def\thetabar{\bar\theta}
\def\Tr{{\rm Tr}}
\def\one{\mbox{1 \kern-.59em {\rm l}}}
 \def\Nh{\hat{N}}

\newcommand{\half}{{\textstyle {1 \over 2}}}

\def\a{\alpha}      \def\da{{\dot\alpha}}
\def\b{\beta}       \def\db{{\dot\beta}}
\def\c{\gamma}  \def\G{\Gamma}  \def\cdt{\dot\gamma}
\def\d{\delta}  \def\D{\Delta}  \def\ddt{\dot\delta}
\def\e{\epsilon}        \def\vare{\varepsilon}
\def\f{\phi}    \def\F{\Phi}    \def\vvf{\f}
\def\h{\eta}
\def\k{\kappa}
\def\l{\lambda} \def\L{\Lambda}
\def\m{\mu} \def\n{\nu}
\def\o{\omega}
\def\p{\pi} \def\P{\Pi}
\def\r{\rho}
\def\s{\sigma}  \def\S{\Sigma}
\def\t{\tau}
\def\th{\theta} \def\Th{\Theta} \def\vth{\vartheta}
\def\X{\Xeta}
\def\z{\zeta}
\def\de{\partial}

\def\cA{{\cal A}} \def\cB{{\cal B}} \def\cC{{\cal C}}
\def\cD{{\cal D}} \def\cE{{\cal E}} \def\cF{{\cal F}}
\def\cG{{\cal G}} \def\cH{{\cal H}} \def\cI{{\cal I}}
\def\cJ{{\cal J}} \def\cK{{\cal K}} \def\cL{{\cal L}}
\def\cM{{\cal M}} \def\cN{{\cal N}} \def\cO{{\cal O}}
\def\cP{{\cal P}} \def\cQ{{\cal Q}} \def\cR{{\cal R}}
\def\cS{{\cal S}} \def\cT{{\cal T}} \def\cU{{\cal U}}
\def\cV{{\cal V}} \def\cW{{\cal W}} \def\cX{{\cal X}}
\def\cY{{\cal Y}} \def\cZ{{\cal Z}}

\def\ua{\underline{\alpha}}
\def\ub{\underline{\phantom{\alpha}}\!\!\!\beta}
\def\uc{\underline{\phantom{\alpha}}\!\!\!\gamma}
\def\um{\underline{\mu}}
\def\ud{\underline\delta}
\def\ue{\underline\epsilon}
\def\una{\underline a}\def\unA{\underline A}
\def\unb{\underline b}\def\unB{\underline B}
\def\unc{\underline c}\def\unC{\underline C}
\def\und{\underline d}\def\unD{\underline D}
\def\une{\underline e}\def\unE{\underline E}
\def\unf{\underline{\phantom{e}}\!\!\!\! f}\def\unF{\underline F}
\def\unm{\underline m}\def\unM{\underline M}
\def\unn{\underline n}\def\unN{\underline N}
\def\unp{\underline{\phantom{a}}\!\!\! p}\def\unP{\underline P}
\def\unq{\underline{\phantom{a}}\!\!\! q}
\def\unQ{\underline{\phantom{A}}\!\!\!\! Q}
\def\unH{\underline{H}}

\def\As {{A \hspace{-6.4pt} \slash}\;}
\def\bs {{b \hspace{-6.4pt} \slash}\;}
\def\Ds {{D \hspace{-6.4pt} \slash}\;}
\def\ds {{\del \hspace{-6.4pt} \slash}\;}
\def\ss {{\s \hspace{-6.4pt} \slash}\;}
\def\ks {{ k \hspace{-6.4pt} \slash}\;}
\def\ps {{p \hspace{-6.4pt} \slash}\;}
\def\pas {{{p_1} \hspace{-6.4pt} \slash}\;}
\def\pbs {{{p_2} \hspace{-6.4pt} \slash}\;}
\def\Ps {{P \hspace{-6.4pt} \slash}\;}
\def\Qs {{Q \hspace{-6.4pt} \slash}\;}

\def\Fh{\hat{F}}
\def\Vh{\hat{V}}
\def\Xh{\hat{X}}
\def\ah{\hat{a}}
\def\xh{\hat{x}}
\def\yh{\hat{y}}
\def\ph{\hat{p}}
\def\xih{\hat{\xi}}
\def\psit{\tilde{\psi}}
\def\Psit{\tilde{\Psi}}
\def\tht{\tilde{\th}}
\def\lt{\tilde{\lambda}}
\def\hl{\hat{\lambda}}
\def\hlt{\hat{\tilde{\lambda}}}
\def\llt{\tilde{l}}
\def\At{\tilde{A}}
\def\Qt{\tilde{Q}}
\def\Rt{\tilde{R}}
\def\Nt{\tilde{N}}

\def\at{\tilde{a}}
\def\st{\tilde{s}}
\def\ft{\tilde{f}}
\def\pt{\tilde{p}}
\def\qt{\tilde{q}}
\def\vt{\tilde{v}}
\def\nt{\tilde{n}}

\def\delb{\bar{\partial}}
\def\bz{\bar{z}}
\def\bD{\bar{D}}
\def\bB{\bar{B}}

\def\bk{{\bf k}}
\def\bl{{\bf l}}
\def\bp{{\bf p}}
\def\bq{{\bf q}}
\def\br{{\bf r}}
\def\bx{{\bf x}}
\def\by{{\bf y}}
\def\bR{{\bf R}}
\def\bV{{\bf V}}

\def\d{\delta}\def\D{\Delta}\def\ddt{\dot\delta}
\def\pa{\partial} \def\del{\partial}
\def\xx{\times}
\def\uno{\mbox{1 \kern-.59em {\rm l}}}
\def\trp{^{\top}}
\def\inv{^{-1}}
\def\dag{{^{\dagger}}}
\def\pr{^{\prime}}
\def\lan{\langle}
\def\ran{\rangle}
\def\rar{\rightarrow}
\def\lar{\leftarrow}
\def\lrar{\leftrightarrow}
\newcommand{\0}{\,\!}      
\def\one{1\!\!1\,\,}
\def\im{\imath}
\def\jm{\jmath}
\newcommand{\tr}{\mbox{tr}}
\newcommand{\slsh}[1]{/ \!\!\!\! #1}
\def\vac{|0\rangle}
\def\lvac{\langle 0|}
\def\hlf{\frac{1}{2}}
\def\ove#1{\frac{1}{#1}}
\def\Box{\square}
\def\ZZ{\mathbb{Z}}
\def\CC#1{({\bf #1})}
\def\bcomment#1{}
\def\bfhat#1{{\bf \hat{#1}}}
\def\VEV#1{\left\langle #1\right\rangle}
\newcommand{\ex}[1]{{\rm e}^{#1}} \def\ii{{\rm i}}
\def\rr{{\rm r}} \def\rs{{\rm s}}\def\rv{{\rm v}}
\def\ri{{\rm i}}\def\rj{{\rm j}}
\newcommand{\lrbrk}[1]{\left(#1\right)}
\newcommand{\sfrac}[2]{{\textstyle\frac{#1}{#2}}}

\def\Li{{\rm Li}_2}

\font\mybb=msbm10 at 12pt
\def\bb#1{\hbox{\mybb#1}}

\font\myBB=msbm10 at 18pt
\def\BB#1{\hbox{\myBB#1}}

\begin{document}

\begin{flushright}
QMUL-PH-08-15
\end{flushright}

\vspace{20pt}

\begin{center}

{\Large \bf A note on dual superconformal symmetry  of   }
\\
\vspace{0.3cm}
{\Large \bf   the $\mathcal{N}=4$ super Yang-Mills S-matrix }
\vspace{11pt}
\vspace{32pt}

{\mbox {\bf Andreas Brandhuber, Paul Heslop and Gabriele Travaglini}}%
\footnote{
{\sffamily \{\tt a.brandhuber, p.j.heslop, g.travaglini\}@qmul.ac.uk }}

{\em Centre for Research in String Theory\\
Department of Physics\\
Queen Mary, University of London\\
Mile End Road, London, E1 4NS\\
United Kingdom
 }

\vspace{30pt} {\bf Abstract}

\end{center}

\noindent
We present a supersymmetric recursion relation 
for tree-level scattering amplitudes in $\cN=4$ super Yang-Mills. 
Using this recursion relation, we prove that the tree-level $S$-matrix 
of the maximally supersymmetric theory is covariant under  
dual superconformal transformations. We further analyse  the consequences 
that the transformation properties of the trees under this symmetry  
have on those of the loops. In particular, we show that the coefficients of the expansion of
generic one-loop amplitudes in a basis of  pseudo-conformally invariant 
scalar box functions transform  covariantly under dual superconformal symmetry, and
in exactly the same way as the corresponding tree-level amplitudes.

\noindent

\setcounter{page}{0}
\thispagestyle{empty}
\newpage


\section{Introduction and background}
\setcounter{footnote}{0}

In an interesting paper \cite{dhks}, Drummond, Henn, Korchemsky, and Sokatchev (DHKS) 
have proposed that scattering amplitudes in planar $\cN=4$ super
Yang-Mills (SYM) theory have a novel superconformal symmetry, termed dual
in order to distinguish it from the ordinary superconformal symmetry. 

This symmetry has also been explained 
very recently from the string theory standpoint \cite{bermal,tse} using a T-duality of the 
superstring theory on $AdS_5 \times S^5$ which involves a bosonic T-duality \cite{am}, 
accompanied by  a new fermionic T-duality. The combined effect of these T-dualities is to 
map the original string sigma model into a dual sigma model identical to the original one.   
The T-duality exchanges the original with the dual superconformal symmetries; 
furthermore,  the strong coupling calculation of the amplitudes  in the dual sigma model 
turns out to be  technically identical to that of a Wilson loop with a special closed contour,  
constructed by gluing together the  momenta of scattered particles  
following the order of the insertions of the string vertex operators \cite{am}.  
Surprisingly, calculations of the same Wilson loops in $\cN=4$ SYM 
at weak coupling at one \cite{dks,bht} and two loops \cite{dhks4,dhks5,dhksbum,dhks6} 
are in perfect  agreement  with the MHV scattering amplitudes of the $\mathcal{N}=4$ theory
calculated in \cite{bddk,abdk,2l5pt,seven}. 
See \cite{cento} for a recent review on the duality between scattering amplitudes and Wilson loops.

According to the proposal put forward in \cite{dhks},  all tree-level superamplitudes 
are covariant under dual superconformal symmetry, and  
their transformations should be precisely the same as those of the 
supersymmetric expression introduced by Nair \cite{Nair} which generalises the usual MHV amplitudes%
\footnote{A superamplitude can be thought of as a generating function that combines all tree 
amplitudes with a fixed
number of external lines and fixed total helicity into one supersymmetric quantity. More details on this
formalism are presented later in this Section and in Section 2.}.
It is one of the goals of this paper  to prove this statement, {\it i.e.}~to show that all tree-level 
superamplitudes of the  $\cN=4$ theory transform covariantly under this symmetry, 
and in exactly the same way as the MHV superamplitude.

In order to achieve this goal, we look for a method  to compute amplitudes which 
respects superconformal covariance at the diagrammatic level. 
We claim that one such method is given by an appropriate supersymmetric extension 
of the BCF recursion relation \cite{bcf, bcfw}, which we will write down explicitly.%
\footnote{An $\cN=4$ supersymmetric recursion relation using  the triple shifts 
of \cite{Risager} has recently been written down in  \cite{Massimo}.  }

The original motivation for this claim  comes from the explicit inspection of the
recursive diagrams for the next-to-MHV (NMHV) split-helicity gluonic amplitudes%
\footnote{Split-helicity amplitudes have all  positive helicity gluons and
all negative helicity gluons adjacent.}
calculated in 
 \cite{bcf,splitbritto}. 
As was observed in \cite{dhks}, all gluonic split-helicity amplitudes are covariant. 
Furthermore, one can easily verify that this covariance is  realised separately in each recursive diagram, 
as a direct inspection of the derivations of \cite{bcf,splitbritto} shows. 
Note, however, that non-split-helicity amplitudes do not transform covariantly \cite{dhks} and they
have to be packaged together with the split-helicity amplitudes into superamplitudes, which according
to \cite{dhks} should transform covariantly in general.
So far this claim has been verified for the case of MHV and NMHV amplitudes.
To extend this observation to a full proof of dual superconformal covariance of the tree-level $S$-matrix
of the $\cN=4$ theory, we will first write down an appropriate supersymmetric recursion relation satisfied by the superamplitudes in the maximally supersymmetric theory. 

In the supersymmetric formalism of \cite{Nair},  to each particle in the $\cN=4$ theory one associates 
the usual commuting spinors
$\l_\a$, $\lt_{\da}$ (in terms of which the momentum of the $i^\mathrm{th}$
particle is $p_{\a \da}^{i} = \l_{\a}^{i} \lt_{\da}^{i}$), as well
as anticommuting variables $\eta^{A}_{i}$, where $A=1,\ldots,4$ is an $SU(4)$ index. 
The supersymmetric amplitude can then be expanded in powers of the
$\cN=4$ superspace coordinates $\eta_{A}^{i}$ for the different particles, 
and each term of this expansion corresponds to a particular scattering
amplitude in $\cN=4$ SYM. A term containing $p$ powers of
$\eta_{A}^{i}$ corresponds to a scattering process where the
$i^\mathrm{th}$  particle has helicity $h_i = 1 - p/2$ \cite{GK}. Explicitly, the $n$-point 
MHV superamplitude is \cite{Nair} 
\beq
\label{NairMHV}
\cA_{\rm MHV} (1, \ldots , n) \ = \ i (2\pi)^4 \, 
{\delta^{(4)} (\sum_{i=1}^{n} \l_i \lt_i ) \, \delta^{(8)} ( \sum_{i=1}^{n} \eta_i \l_i ) 
\over \lan12\ran \cdots \lan n1\ran } 
\ , 
\eeq
where, as usual, $\lan i  j\ran := \eps_{\a \b} \l_i^\a \l_j^\b$.  

The dual superconformal symmetry becomes more transparent after introducing appropriate dual coordinates \cite{dhks}. These turn out to be 't Hooft's region (or T-dual) momenta 
\beq
p_{i, \a \dot{\a}} \ = \ (x_{i}- x_{i+1})_{\a \dot{\a}}
\ , 
\eeq
along with their supersymmetric partners $\theta_i^{A \a}$ introduced in 
\cite{dhks} as
\beq
\eta_i^A \l_{i}^{\a} \ =  \ \theta_i^{A \a} - \theta_{i+1}^{A \a}
\ . 
\eeq
It is important to note that these coordinates are appropriate for characterising planar diagrams only, where one can express the momentum carried by one line as the difference of the momenta 
of the two regions of the plane separated by the line. The dual
momenta also play an important r\^{o}le 
in  the discussion of pseudo-conformal properties of integral functions in \cite{magic}.  

The dual momenta $x_i$, $i=1, \ldots , n$, are such that the momenta of each particle are null, 
{\it i.e.} $(x_i - x_{i+1})^2 = 0$, and momentum conservation becomes automatic in this 
formalism.  It therefore makes perfect sense to act with inversions on the dual momenta,  
which transform as%
\footnote{Special conformal transformations are obtained as an inversion followed by a translation, and a 
further inversion. Combining this with supersymmetry transformations,
one generates all the superconformal transformations. Since the dual
supersymmetries are either manifest or are related to ordinary special superconformal symmetries~\cite{dhks},
which obviously are symmetries of tree-level  $\cN$=4 SYM, 
invariance of the $S$-matrix under the full dual
superconformal symmetry requires only showing invariance under dual
inversions.
}

\beq
\label{xtran}
x_{i, \a \dot{\b}}  \ \to \ {x_{i, \b \dot{\a}} \over x_i^2} 
\ .  
\eeq
Similarly, the differences of fermionic variables $\theta_i$ of adjacent particles are constrained to be {\it on shell}, namely
\beq
(\theta_i - \theta_{i+1} ) \lambda_i \ = 0 
 \ , 
 \eeq
and the $\theta_i$ transform under inversions as \cite{dhks} 
\beq
\label{thtran}
\theta_i^{A \a} \ \to \ (x_i^{-1})^{\dot{\a} \b} \th_{ i, \b}^{A}
\ . 
\eeq
For completeness, we also present the transformation of the variables
$\eta^A$ which can be deduced from the transformation above~\cite{dhks},
\begin{equation}\label{inveta}
    I[\eta^A_i] = {\frac{x^2_{i}}{x^2_{i+1}}}\ \left(\eta^A_i -
      {\theta^A_i x^{-1}_i\tilde\lambda_i} \right)\, .
\end{equation}

Using these transformations,  it is easy to see that  the MHV
superamplitude~\eqref{NairMHV} transforms covariantly under inversions,
\begin{align}\label{MHVtrans}
  \cA_{\rm MHV}(1,2,\dots,n) \rightarrow \cA_{\rm MHV}(1,2,\dots,n) \  \prod_{k=1}^n
  {x_k^2} \ .
\end{align}
After introducing a supersymmetric version of BCF on-shell recursion relations, 
we will show that this transformation property \eqref{MHVtrans} 
is maintained for any tree-level superamplitude in 
$\mathcal{N}=4$ SYM.

After this short discussion of dual superconformal properties of tree-level amplitudes, 
we now move on to consider loop amplitudes. 
There the situation is more subtle due to the appearance of infrared divergences in the 
scattering amplitudes, which manifest themselves as ultraviolet divergences in the dual
Wilson loops, due to the presence of cusps in the contour.   
Interestingly, it was shown in \cite{dhks4,dhks5} that by performing 
dual conformal transformations on the lightlike Wilson loops in the $\cN=4$ theory 
one can derive anomalous Ward identities, which turn out to be consistent with 
the BDS ansatz \cite{bds} for the 
exponentiated form of the $n$-point MHV scattering amplitude of the $\cN=4$ theory. 
In the four- and five-point case, the solution to the Ward identity is actually unique
up to a finite constant, whereas for $n\geq6$  particles, there is room for a conformally invariant 
discrepancy function, compared to the BDS ansatz,  
which was indeed found to be nonzero in \cite{dhks6,seven}. 
In this paper we focus  our attention on the coefficients of the expansion of one-loop amplitudes in 
$\cN=4$ SYM in terms of integral box functions, and to their transformation properties under  
dual superconformal transformations. We  will show that these coefficients
are covariant under superconformal symmetry and exhibit the same transformation properties 
as those of the tree-level superamplitudes.  
The main tool in this analysis is the use of quadruple cuts of \cite{bcfgen} which, 
crucially, can be performed  in four dimensions, since all one-loop amplitudes of 
the maximally supersymmetric theory are four-dimensional cut constructible \cite{fusing}. 
This simplifies the analysis considerably, by-passing  dimensional regularisation.
As an added bonus of this analysis, we will obtain an independent proof of the covariance of the 
tree-level superamplitudes.

The rest of the paper is organised as follows: in Section 2 we introduce a supersymmetric generalisation of the BCF recursion relations, present the $\overline{\mathrm{MHV}}$ three-point superamplitude
and discuss the behaviour of superamplitudes under large complex deformations (shifts). 
In Section 3 we give some simple applications of the supersymmetric recursion relations. Readers who are familiar
with the formalism may wish to skip this part. In Section 4 we use the supersymmetric recursion
relations developed in Section 2 and 3 to prove that all tree-level superamplitudes in 
$\cN=4$ SYM transform uniformly under dual conformal transformations. 
Finally, in Section 5 we prove that the coefficients that appear in the expansion of generic one-loop superamplitudes in $\cN=4$ SYM in a basis of scalar box functions transform covariantly in exactly  the same way as the corresponding tree-level  amplitudes.

\section{$\mathcal{N}=4$ supersymmetric recursion relations}

In this section we write down a supersymmetric recursion relation using  two-particle shifts.%
\footnote{As mentioned earlier, an $\mathcal{N}=4$ supersymmetric recursion relation was written down in \cite{Massimo}  for NMHV amplitudes using a set of three antiholomorphic shifts suggested 
by Risager \cite{Risager}. 
In that case it is immediate to see that the two amplitudes appearing in the corresponding recursion relation must have the MHV helicity configuration. Indeed, the corresponding diagrams are the super MHV diagrams 
considered in Section 5 of \cite{GGK}.}
These shifts can be nicely formulated using the dual superspace variables 
introduced  in~\cite{dhks}. 
The recursion relation using conventional two-particle shifts requires
the three-point anti-MHV amplitude as well as the MHV amplitude as input. We
will thus require a three-point anti-MHV superamplitude and 
we propose precisely such a superamplitude in the next subsection.
We then address the important issue of the large-$z$ behaviour of the $\cN=4$ superamplitudes  
in subsection \ref{sec:large-z-behaviour}, where we prove that the superamplitude calculated 
with the supersymmetric recursion relation agrees with that obtained by standard methods.

In order to set up the formalism, we briefly review the derivation of the BCF recursion relations. 
The key property entering these recursion relations is  
factorisation on  multi-particle poles (or collinear factorisation, for MHV amplitudes). 
To exploit this efficiently, one considers a  particular deformation of an amplitude which
shifts the spinors of two of the $n$ massless external particles,
labelled here as $i$ and $j$, as \cite{bcfw} 
\beq
\label{eq:bcfwshifts}
\widetilde{\lambda}_i\ \to\ \hat{\widetilde{\lambda}}_i \ := \ \widetilde{\lambda}_i+z\widetilde{\lambda}_j
\ ,
\qquad 
\lambda_j\ \to \ \hat{\lambda}_j \ := \ \lambda_j-z\lambda_i
\ , 
\eeq
where $z$ is the complex parameter characterising the deformation. 
The spinors $\lambda_i$ and $\widetilde{\lambda}_j$ are left unshifted. 
The deformations \eqref{eq:bcfwshifts} are chosen in such a way that 
the corresponding shifted momenta 
\beq
\label{shiftmom}
\hat{p}_i (z) := \l_i \hat{\widetilde{\l}}_i\, = \, p_i + z \l_i \widetilde{\l}_ j \ , \quad 
\hat{p}_j (z) := \hat{\l}_j \widetilde{\l}_j\,  = \, p_j  - z \l_i \widetilde{\l}_ j
\ , 
\eeq
are on shell for all complex $z$. Furthermore, 
$p_i(z)+p_j(z)=p_i + p_j$.
Hence the quantity $\cA(p_1,\ldots,p_i(z),\ldots,p_j(z),\ldots,p_n)$
is a well-defined one complex parameter  family of scattering amplitudes, 
parametrised  by $z$.

One then considers the following  contour integral, where the contour $\cC$ is the circle at infinity in
the complex $z$-plane,
\begin{equation}
\label{van}
\frac{1}{2\pi i} \oint_{\cC} \!dz \frac{\cA(z)}{z}
\ .
\end{equation}
The integral in \eqref{van} vanishes if $\cA(z) \to 0$ as $z \to
\infty$~\footnote{We prove this property for a large portion of the
  superamplitude in Section~\ref{sec:large-z-behaviour} and use supersymmetry
  to argue that this is enough to determine the entire superamplitude.}. It  then follows from Cauchy's  theorem
that we can write the amplitude we wish to calculate, $\cA(0)$,  as a sum of residues
of $\cA(z)/z$, 
\begin{equation}
\cA(0)\ =\ -\!\!\!\!\!\!\!\!\!\!\sum_{\substack{\textrm{poles of $\cA(z)/z$}\\\textrm{excluding $z\!=\!0$}}}
\!\!\!\!\!\!\!\textrm{Res} \left[\frac{\cA(z)}{z}\right]
\ .
\end{equation}
At tree level,  $\cA(z)$ has only simple poles in $z$.
A pole at $z\!=\!z_{P}$ is associated with a shifted momentum
$\hat{P} := P(z_P)$ flowing through an internal propagator becoming null.
The residue at this pole is then obtained by factorising the shifted amplitude on
this pole. The result is that 
\beq
\label{ampli}
\cA \ = \ 
\sum_{P} 
\sum_h \cA_L^h(z_{P})
 \frac{i}{P^2} \cA_R^{-h}(z_{P})
 \ ,
\end{equation}
where the sum is over the possible  assignments of the helicity $h$ of the intermediate state, and over 
all possible $P$ such that precisely one of the shifted momenta, say $\hat{p}_i$,  is contained in $P$. 

The left and right hand amplitudes $\cA_L$ and $\cA_R$ are 
well-defined amplitudes only for $z\!=\!z_{P}$,  when
$P(z)$ becomes  null.  We call $\lambda_{\hat{P}}$ and $\lt_{\hat{P}}$ the spinors associated to the 
internal,  on-shell momentum  $\hat{P}$, so that  
$\hat{P} := \lambda_{\hat{P}} \lt_{\hat{P}}$. 
Notice that the intermediate propagator is evaluated with unshifted kinematics.

Since a momentum invariant involving both (or neither) of the shifted legs $i$
and $j$ does not give rise to a pole in $z$, the shifted legs $i$ and
$j$ must always appear on opposite sides of the factorisation channel.
In order to limit the number of recursive diagrams, it is very
convenient to shift adjacent legs. 
In this case, the sum over $P$ in \eqref{ampli} is just a single sum. 
In the following we will do this, so that the shifted legs will always be $i$ and $j=i+1$. 
We will denote the shift in \eqref{eq:bcfwshifts} with the standard notation 
$ [i\,  i+1\ran$. 

Now for the supersymmetric version of the BCF recursion relation. 
Firstly, we  notice that 
it is very  easy to describe the  shifts \eqref{eq:bcfwshifts}, \eqref{shiftmom}
using dual (or region) momenta. One simply defines
\beq
\hat{p} _i \ := \ x_i - \hat{x}_{i+1} \ , 
\qquad \hat{p}_{i+1} \ := \ \hat{x}_{i+1} - x_{i+2} 
 \ ,
 \eeq
where we have introduced a shifted region momentum 
\beq
\hat{x}_{i+1} \ := \ x_{i+1} - z \, \lambda_i \widetilde{\l}_{i+1}
 \ . 
 \eeq
 Notice that this is the  only  region momentum that is  affected by the shifts%
 \footnote{This is true only if adjacent legs are shifted. If $i$ and $j$ are not adjacent, then
 region momenta $x_{i+1} \ldots x_j$ are all shifted by $- z \, \lambda_i \widetilde{\l}_{j}$.}.
Therefore in the supersymmetric case we expect that 
$\theta_{i+1}$ is shifted but all other $\theta$'s  remain unshifted.
This implies that 
\beq
 \theta_i-\theta_{i+2}  =  \ \eta_i \l_i + \eta_{i+1} \l_{i+1} 
 \ , 
 \eeq
should remain unshifted. This is in complete similarity to the fact
that the sum of the shifted momenta is unshifted,  $\hat{p}_i +
\hat{p}_{i+1} = p_i + p_{i+1}$.   
Now, in the case of the $[i\,  i+1 \ran$ shift employed here, we have
shifted $\lambda_{i+1}$ according to~(\ref{eq:bcfwshifts}) and so we
can achieve this 
by shifting $\eta_i$ to 
\beq 
\hat{\eta}_i = \eta_i + z \, \eta_{i+1} 
\ ,
\eeq
and leaving $\eta_{i+1}$ unshifted.
This then gives the shifted $\theta_{i+1}$
\beq
\label{st}
\hat{\theta}_{i+1}    \ := \ \theta_{i+1} - z \, \eta_{i+1} \l_i 
\ . 
\eeq

The recursion relation builds up tree-level amplitudes recursively
from lower point amplitudes. The starting point of this process is
the MHV superamplitude~\eqref{NairMHV}
(in fact just the three-point MHV superamplitude  is needed) together
with the three-point anti-MHV superamplitude which we present and discuss  in the next
section. 

The supersymmetric recursion relation follows  from arguments similar to those which 
led to~(\ref{ampli}). We have 
\beq
\label{superampli}
\cA \ = \ 
\sum_{P} 
\int\!\!d^4\eta_{\hat{P}} \  \cA_L(z_{P})
 \frac{i}{P^2} \cA_R(z_{P})
 \ , 
\end{equation}
where $\eta_{\hat{P}}$ is the anticommuting  variable associated to the internal, 
on-shell leg with momentum $\hat{P}$.  

Note that in the case of superamplitudes it does not make sense to assign individual
helicities to the external particles, and every superamplitude is
characterised by the number of external particles and 
its total helicity, which is the sum of the helicities of all external particles. 
In the recursion relation (\ref{superampli})
we have an important constraint on  $\cA_L$ and  $\cA_R$, namely the
total helicity of $\cA_L$ plus the total helicity of $\cA_R$ must equal the total helicity of the
full amplitude $\cA$. This condition replaces the sum over internal helicities in the
standard BCF recursion~(\ref{ampli}).

\subsection{ Supersymmetric anti-MHV three-point amplitudes}
In writing down recursion relations, 
one  needs as a starting point the three-point MHV and $\overline{\mathrm{MHV}}$ amplitudes.
Whereas the former are given by the usual Nair formula, we also
require a supersymmetric expression for the latter. 
We claim that this is 
\beq
\label{3ptmhvbar}
\cA_{\overline{\mathrm{MHV}}} (1, 2, 3) \ = \ i (2 \pi)^4 \, 
{\d^{(4)} (p_1 + p_2 + p_3 ) \delta^{(4)} ( \eta_1 [23] \, + \, \eta_2 [ 31] \, + \, \eta_3 [12] ) \over [12] \, [23]\, [31] } 
\ . 
\eeq
For example, the gluonic amplitude $\cA (1_g^+, 2_g^+ , 3_g^-) = [12]^3 / ( [23][31]) $ is immediately obtained by extracting the 
component $\prod_{A=1}^4 \eta_3^A$ of \eqref{3ptmhvbar}. 

In order to verify that \eqref{3ptmhvbar} is supersymmetric, we multiply it by 
the sum of the supercharges $\sum_{i=1}^3 Q_{i; \a}^A := \sum_{i=1}^{3} 
\eta_{i }^A\l_{i, \a}$. 
Upon acting on the combination of delta functions in \eqref{3ptmhvbar}, 
one has 
\beqa
\sum_{i=1}^3 Q_{i; \a}^A &\to&  {-\eta_2^A [31] - \eta_3^A [12] \over [23] } \l_{1, \a} 
 \ + \ \eta_2^A \lambda_{2,\alpha} \ + \ \eta_3^A
 \lambda_{3,\alpha}\nonumber \\
&=& \eta_2^A \, {\l_{2,\a} [23] + \l_{1,\a} [13] \over [23] } + 
\eta_3^A \, {\l_{3, \a} [23] + \l_{1, \a} [21] \over [23] } 
\nonumber \\ 
&=& 0 
\ , 
\eeqa
where the last equality follows from momentum conservation
$\l_1 \lt_1 + \l_2 \lt_2 + \l_3 \lt_3 = 0$. 
As discussed in~\cite{dhks}, the condition for the amplitude to be
invariant  under the second set of supersymmetry generators
is
\begin{align}
\bar Q_{A\dot \alpha}\,  \cA_{\overline{\mathrm{MHV}}}(1,2,3)  \, = \,  
\sum_{i=1}^3 \tilde \lambda_{i \dot \alpha}
{\partial \over \partial 
    \eta_i^A} \cA_{\overline{\mathrm{MHV}}}(1,2,3) \, = \, 0\ .
\end{align}
If we act with the operator $\bar Q_{A\dot \alpha}$ on the argument of the fermionic delta function
in~\eqref{3ptmhvbar}, we obtain 
\begin{align}
\bar Q  (\eta_1 [23] \, + \, \eta_2 [ 31] \, + \, \eta_3 [12] )  =&\ \tilde\lambda_1[23]+\tilde\lambda_2[31]+\tilde\lambda_3[12]\ = \ 0\ , 
\end{align}
thus proving that $\cA_{\overline{\mathrm{MHV}}}$ is  invariant also under the 
$\bar Q$ supersymmetries.

Next we would like to show explicitly that \eqref{3ptmhvbar} transforms as a three-point amplitude, {\it i.e.}
that 
\beq
\cA_{\overline{\mathrm{MHV}}} (1, 2, 3) \ \to  x_1^2 x_2^2 x_3^2 \ \cA_{\overline{\mathrm{MHV}}} (1, 2, 3)
\ , 
\eeq
under a conformal inversion. This is slightly nontrivial due to the absence of the usual  eight-dimensional  
delta function of supermomentum conservation in \eqref{3ptmhvbar}. 

The proof is very simple. Firstly, we notice that since 
\beq\label{rj}
{1\over [12][23][31]} \ \to \  {x_1^2 x_2^2 x_3^2 \over [12][23][31]} \ , 
\eeq
we have to show that the combination 
$\delta^{(4)} ( p_1 + p_2 + p_3 ) \delta^{(4)} ( \eta_1 [23] \, + \, \eta_2 [ 31] \, + \, \eta_3 [12] )$ 
is invariant under inversions. 

In order to see this, we recall  that 
\beq
\label{orfeo}
 \delta^{(4)} ( \eta_1 [23] \, + \, \eta_2 [ 31] \, + \, \eta_3 [12] ) := 
 \prod_{A=1}^4 ( \eta_1^A [23] \, + \, \eta_2^A [ 31] \, + \, \eta_3^A [12] )
\ . 
\eeq
Multiplying and dividing by $\l_{1, \alpha}$ for a fixed $\alpha$, one gets
\beq
\label{euridice}
( \eta_1^A [23] \, + \, \eta_2^A [ 31] \, + \, \eta_3^A [12] ) \l_{1, \alpha} \ = \ [23] \, (\theta_1 - \theta_4)^A_{\alpha} 
\ 
\eeq
(notice that we have broken the cyclicity  of the $\theta$ variables). 
Hence we can write 
\beq
\label{bio} 
 \delta^{(4)} ( \eta_1 [23] \, + \, \eta_2 [ 31] \, + \, \eta_3 [12] ) \ = \ 
\left( { [23] \over \lambda_{1, \a}} \right)^4 
\, \prod_{A=1}^{4} (\theta_1 - \theta_4)^A_{\alpha} 
\ , 
\eeq
at fixed (and arbitrary) $\alpha$. The transformation properties of \eqref{bio} are manifest, using $\l_1 \to x_1^{-1} \l_1$, 
$[23] \to [23] / x_1^2$, and $\theta_1 \to x_1^{-1} \theta_1$,  $\theta_4 \to x_4^{-1} \theta_4 = 
x_1^{-1} \theta_4$, where the last step follows since the expression \eqref{3ptmhvbar} contains a  
$\delta^{(4)} (p_1 + p_2 + p_3 ) = \delta^{(4)} (x_1 - x_4)$. 
Therefore
\beqa
\label{bio2} 
\left( { [23] \over \lambda_{1, \a}} \right)^4 
\, \prod_{A=1}^{4} (\theta_1 - \theta_4)^A_{\alpha}
& \to  &  
{1 \over  (x_1^2)^4}  \, 
\left( { [23] \over (x_1^{-1})^{\dot\alpha \beta} \lambda_{1, \beta}} \right)^4 \, 
\prod_{A=1}^{4} (x_1^{-1})^{\dot\alpha \beta} (\theta_1 - \theta_4)^A_{\beta}
\nonumber \\ 
&=& {1 \over  (x_1^2)^4} \, \prod_{A=1}^4 ( \eta_1^A [23] \, + \, \eta_2^A [ 31] \, + \, \eta_3^A [12] )
\ , 
\eeqa
where the last equality follows in a way completely similar to that used to derive \eqref{bio}, 
except that one multiplies and divides by $x_1^{-1} \l_1$. 
Finally, comparing \eqref{bio2} and  \eqref{bio}, we see that 
\beq
\delta^{(4)}  ( \eta_1 [23] \, + \, \eta_2 [ 31] \, + \, \eta_3 [12] ) \ \to
\left( {1\over x_1^2}\right)^4\, \delta^{(4)}  ( \eta_1[23] \, + \, \eta_2 [ 31] \, + \, \eta_3[12] ) 
\ , 
\eeq
under conformal inversions. Since $\delta^{(4)} (x_1 - x_4) \to (x_1^2)^4\, \delta^{(4)} (x_1 - x_4) $, 
it follows that the combination $\delta^{(4)} ( p_1 + p_2 + p_3 ) \delta^{(4)} ( \eta_1 [23] \, + \, \eta_2 [ 31] \, + \, \eta_3 [12] )$ is invariant, and hence the three-point $\overline{\rm MHV}$ amplitude \eqref{3ptmhvbar}
 transforms correctly as \eqref{rj} under inversions.

To conclude this section, we notice that an expression for the three-point $\overline{\rm MHV}$
has been presented in \cite{ah} which reads%
\footnote{We thank Johannes Henn for bringing this to our attention.}
\beq
\label{nah}
\cA_{\overline{\mathrm{MHV}}}(1,2,3)  =  i (2\pi)^4 \, { \delta^{(4)} (p_1 + p_2 + p_3)
\over [12]\, [23]\, [31]}
\int\! \prod_{i=1}^{3} d^4 \bar{\eta}_i \ e^{ \sum_{i=1}^{3} {\bar{\eta}}_{i; A} \eta_i^A}\ 
\delta^{(8)} ( \bar{\eta}_1 \lt_1 +   \bar{\eta}_2 \lt_2 +\bar{\eta}_3 \lt_3 ) . 
\eeq
It is very easy to perform the $\bar{\eta}$ integrations, and check that \eqref{nah} coincides with our form \eqref{3ptmhvbar}
of the three-point $\overline{\rm MHV}$ superamplitude. 

In Section 3 and 4 we will use \eqref{3ptmhvbar} in specific examples in order to show 
how the supersymmetric recursions and the dual momentum superspace formalism work
in practice.

\subsection{Large $z$ behaviour of the supersymmetric amplitudes
  ${\cA}(z)$}
\label{sec:large-z-behaviour}
In the remainder of this section we want to discuss a crucial ingredient in the derivation of the supersymmetric
recursion formula~\eqref{superampli}. The argument leading to~(\ref{ampli}) and its supersymmetric version~(\ref{superampli}) requires that the $z$-shifted amplitude
vanishes as%
\footnote{The large-$z$ behaviour of amplitudes in $\cN=4$ was also addressed in 
\cite{Luo:2005rx} and in \cite{Massimo}, and in the very recent paper \cite{cahk}.} 
$z\rightarrow \infty$. In the case of component gluon amplitudes, this
issue was addressed in \cite{bcfw} using MHV diagrams, as well as Feynman diagrams. 
There,  it was shown that when the two gluons associated with the 
shifted momenta (recall we are using $[i\,  i+1\ran$ shifts) have positive helicity, the amplitude vanishes as 
$z\rightarrow \infty$.

When translated to the supersymmetric
case,  this argument implies that the $z$-shifted superamplitude
$\cA(\eta_i=\eta_{i+1}=0; z)\rightarrow 0$ as $z\rightarrow \infty$. 
The BCFW argument then states that the recursion relation is valid for
$\eta_i=\eta_{i+1}=0$. In other words, defining the function
\begin{align}\label{fdef}
f:=  \cA_{\rm{recursion}} - \cA
  \ , 
\end{align}
where by $\cA_{\rm{recursion}}$ we denote the result of
performing the calculation using the supersymmetric recursion
formula~\eqref{superampli}, and $\cA$ is the correct superamplitude, we have
that the function $f$ vanishes whenever $\eta_i=\eta_{i+1}=0$.

Here, instead of showing directly that the complete
superamplitude vanishes at large $z$, we argue directly, using
supersymmetry, that the recursion relation does give the
correct full superamplitude, given that we know they agree for $\eta_i=\eta_{i+1}=0$.
In order to do this, we make use of  $\bar{Q}$ supersymmetry 
(where $\bar{Q}_{A\dot{\a}}:=\sum_{l=1}^n \tilde{\lambda}_{l \dot
    \alpha}\,  {\partial /  \partial \eta_{l}^{ A}}$),   
which constrains the form of both the amplitude $\cA$ and the result of the
recursion relation $\cA_{\rm{recursion}}$
\footnote{That the recursion relation maintains the $\bar{Q}$ supersymmetry can
be straightforwardly checked. Applying $\bar{Q}$ on a generic recursive diagram 
entering \eqref{superampli} produces two terms, one where
$\bar{Q}$ acts on  $\cA_L$ and one where  $\bar{Q}$ acts on
$\cA_R$. Noting that the $z$-shift  leaves the expression of $\bar{Q}$ unaffected,  
and because of the invariance of  $\cA_L$ and $\cA_R$ under  
$\bar{Q}$ supersymmetry, these two terms combine into a contribution proportional  
to  $ \tilde{\lambda}_{\hat{P}} \int\!d^4\eta_{\hat{P}}\,  \partial/\partial \eta_{\hat{P}} (\cA_L \cA_R)$. 
This is a total derivative,  and hence it  vanishes. 
Therefore  each recursive diagram (and hence the recursion relation)  maintains the 
$\bar Q$ supersymmetry.
The invariance under the $Q$ supersymmetry is  manifest  because of the presence 
of an overall delta function of supermomentum conservation.}. Hence
the difference function $f$ is $\bar Q$ supersymmetric,
\begin{equation}
\label{qbar}
  \bar Q_{A \dot \alpha}   f \ = \  0
  \ .
\end{equation} 
We also notice that $\bar Q$ supersymmetry has been efficiently used in \cite{cahk} 
to show that superamplitudes in $\cN=4$ SYM ($\cN=8$ supergravity) fall of  
as $1/z$ $(1/z^2)$ as $z\to \infty$. 

In order to exploit  the consequences of $\bar Q$ supersymmetry, 
we evaluate~\eqref{qbar} at $\eta_{i}=\eta_{i+1}=0$, and use
the fact that $f$ vanishes when $\eta_i=\eta_{i+1}=0$, to get
\begin{equation}
\label{or-ve}
\left.\left(\tilde{\lambda}_{i \dot
    \alpha}\,  {\partial \over  \partial \eta_{i}^{ A}}+
\tilde{\lambda}_{i+1\, \dot
    \alpha}\,  {\partial \over  \partial \eta_{i+1}^{ A}}\right)  f\right|_{\eta_i=\eta_{i+1}=0} 
    \ =\ 
    0\ .
\end{equation}
For each $A$, \eqref{or-ve} gives  two equations 
which imply that  ${\partial f /  \partial \eta_{i}}={\partial  f /  \partial
  \eta_{i+1}} =0$ 
when $\eta_i=\eta_{i+1}=0$. 
Since $\bar{Q}$ commutes with all 
${\partial /  \partial
      \eta_{l}}$ derivatives,  we can repeat  the above argument 
      for $ {\partial f /  \partial
      \eta_{i}}$ and ${\partial f  /  \partial
      \eta_{i+1}}$ to show that all second derivatives of $f$ with
    respect to $\eta_i, \eta_{i+1}$ also vanish  when $\eta_i=\eta_{i+1}=0$. 
    Continued repetition
    of this argument shows that $f$, and all its partial derivatives with
    respect to $\eta_i$ and $\eta_{i+1}$,
    vanish  when  $\eta_i=\eta_{i+1}=0$,  
    and hence $f$ must  vanish everywhere.

We conclude that  the recursion formula agrees  with the
superamplitude for all $\eta$. 
Several, non-trivial checks of this statement can be found in the next section.

\section{Examples} 

In this section we present some simple applications of the supersymmetric recursion relation.
 
 \subsection{First example, supersymmetric  MHV amplitudes}
 The first example
 is the  case of the MHV amplitude. Here we describe in detail the four-point
 case, but the generalisation to higher numbers of points is straightforward as explained below.
 
 We choose a $[1\, 2\ran$ shift, {\it i.e. }
 \beq
 \label{s1}
\hlt_1 \ = \ \lt_1 + z \lt_2 \ , \quad \hl_2 \ = \ \l_2 - z \l_1 
 \ .  
\eeq
Correspondingly, 
\beq
\hat{p}_1 = x_1  - \hat{x}_2 \ , \quad 
\hat{\eta}_1 \l_1 = \th_1 - \hat{\th}_2 \ , 
\eeq
 where 
\beq 
\hat{x}_2  = x_2  - z \l_1 \lt_2 \ , \quad \hat{\th}_2 = \th_2  - z \eta_2 \l_1 
\ . 
\eeq
Notice that $\hat{\eta}_1  = \eta_1 + z \eta_2$. 
Also, 
\beq
\eta_2 \hl_2 = \hat{\th}_2 - \th_3 \ . 
\eeq
We begin by considering the very simple four-point case. 
The two amplitudes on the left and on the right must be MHV and $\overline{\mathrm{MHV}}$. 
Choosing a $[12\ran$ shift selects the left hand amplitude to be MHV, and the right hand amplitude  to be $\overline{\mathrm{MHV}}$, 
\beqa
\cA_L & = & 
{
\delta^{(4)} ( \hat{1} + 4 + \hat{P} ) \, 
\delta^{(8)} ( \hat{\eta}_1 \l_1 + \eta_4 \l_4 + \eta_{\hat{P}} \l_{\hat{P}}  ) 
\over 
\lan 1 \hat{P} \ran \lan \hat{P}  4 \ran \lan 4 1 \ran 
} 
\ , 
\\ \nonumber
\cA_R & = &  {\delta^{(4)} ( \hat{2} + 3 - \hat{P} ) \, \delta^{(4)} ( \eta_{\hat{P}} [23] + \eta_2 
[3 \hat{P}] + \eta_3 [\hat{P} 2]   ) \over
[\hat{P}2] [23] [3\hat{P}]
} 
\ .  
\eeqa

\begin{figure}
\begin{center}
\scalebox{0.60}{
\fcolorbox{white}{white}{
  \begin{picture}(294,328) (101,-64)
    \SetWidth{0.5}
    \SetColor{CornflowerBlue}
    \Vertex(153,101){46.69}
    \Vertex(347,101){44.55}
    \SetColor{Black}
    \ArrowLine(153,149)(153,224)
    \ArrowLine(345,147)(346,223)
    \ArrowLine(201,103)(301,103)
    \ArrowLine(154,53)(153,-25)
    \ArrowLine(348,55)(349,-16)
    \Text(153,248)[lb]{\Large{\Black{$\hat{1}$}}}
    \Text(345,246)[lb]{\Large{\Black{$\hat{2}$}}}
    \Text(347,-64)[lb]{\Large{\Black{$3$}}}
    \Text(148,-64)[lb]{\Large{\Black{$4$}}}
    \Text(248,141)[lb]{\Large{\Black{$\hat{P}$}}}
   \end{picture}
}
}
\end{center}
\caption{\it Recursive diagram for the {\rm MHV} four-point amplitude. Given the $[12\ran$ shift we have chosen, 
the amplitude on the left must be $\mathrm{MHV}$, and that on the right $\overline{\mathrm{MHV}}$. }
\end{figure}
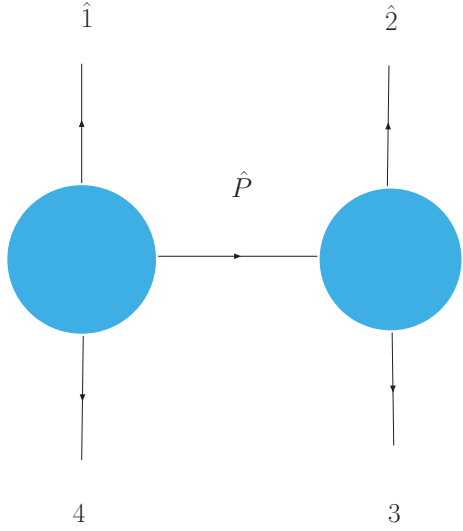

Here we have used the $n$-point MHV superamplitude \eqref{NairMHV}, 
and the expression for the three-point $\overline{\mathrm{MHV}}$ amplitude in \eqref{3ptmhvbar}.

It is easy to see that 
\beqa 
&& \delta^{(8)} ( \hat{\eta}_1 \l_1 + \eta_4 \l_4 + \eta_{\hat{P}} \l_{\hat{P}}  ) \, 
\delta^{(4)} ( \eta_{\hat{P}} [23] + \eta_2 
[3 \hat{P}] + \eta_3 [\hat{P} 2]   ) 
\nonumber \\ 
&=& 
\delta^{(8)} \Big( \sum_{i\in L, R} \eta_i \l_i \Big) \, 
\delta^{(4)} ( \eta_{\hat{P}} [23] + \eta_2 
[3 \hat{P}] + \eta_3 [\hat{P} 2]   )\ , 
\eeqa
so that the amplitude can be written as 
\beqa
\cA(1, 2, 3, 4) &=& 
i\, \delta^{(4)} \Big(\sum_{i\in L,R} p_i \Big) \,
\delta^{(8)} \Big( \sum_{i\in L, R} \eta_i \l_i \Big)  \, A(1,2,3,4)\ , 
\eeqa
where 
\beq
A \ = \ 
{1\over P_{23}^2} \, {1\over \lan 41\ran  [23] \, 
\lan 1 \hat{P}\ran \lan \hat{P} 4\ran [ \hat{P} 2 ] [ 3 \hat{P}] 
} \, 
\int\!d^4\eta_{\hat{P}} \, 
\delta^{(4)} ( \eta_{\hat{P}} [23] + \eta_2 
[3 \hat{P}] + \eta_3 [\hat{P} 2]   )\ . 
\eeq
Completely standard manipulations lead to 
\beq
\lan 1 \hat{P}\ran \lan \hat{P} 4\ran [ \hat{P} 2 ] [ 3 \hat{P}] \ = \ 
\lan12\ran \lan34\ran [23]^2
\ , 
\eeq
hence
\beq
A (1,2,3,4) \ = \  { 1 \over \lan 1 2\ran \lan  23\ran\lan 34\ran\lan 41\ran}
\ . 
\eeq
Hence we reproduce the expected supersymmetric MHV superamplitude. 
Finally, we notice that the recursion relation for  an $n$-point MHV superamplitude 
is a simple generalisation of  that presented above. The only difference is that 
the amplitude on the left hand side of Figure 1 will be an $(n-2)$-point MHV superamplitude. 
The algebra is  identical to that of the four-point example discussed above and leads to the
expected result \eqref{NairMHV}. 

Before moving on to consider five-point amplitudes, we would like to make a comment on the 
the large-$z$ behaviour of the amplitude.  On general grounds, it is known that 
a two-particle shift where the holomorphic spinor associated to a negative  helicity gluon, and the 
antiholomorphic spinor of a positive helicity gluon are shifted leads in general to a bad large-$z$ behaviour  
of the shifted amplitude \cite{bcfw}, {\it i.e.}~the shifted amplitude $A (z)$ does not vanish as $z \to \infty$. 
For example, performing such shifts in the gluonic Parke-Taylor formula may lead to 
a $\cO(z^2)$ growth at large $z$. 
The interesting fact we wish to point out is that  the supersymmetric recursion relation for the MHV superamplitude discussed here  is blind to such bad shifts, as the helicities of the particles in the two superamplitudes entering the recursion relations are not specified, and the recursion relation produces the correct result. Note that this is a general property of the $\cN=4$ supersymmetric recursion relations.

\subsection{Second example, five-point $\overline{\mathrm{MHV}}$ amplitudes}
We continue using the same shifts as in \eqref{s1}. The difference with the previous case is that now the two amplitudes on the left and right hand side of the propagator will both be MHV superamplitudes.
\begin{figure}
\begin{center}
\scalebox{0.60}
{
\fcolorbox{white}{white}{
  \begin{picture}(393,331) (101,-64)
    \SetWidth{0.5}
    \SetColor{CornflowerBlue}
    \Vertex(153,104){46.69}
    \Vertex(347,104){44.55}
    \SetColor{Black}
    \ArrowLine(153,152)(153,227)
    \ArrowLine(345,150)(346,226)
    \ArrowLine(201,106)(301,106)
    \ArrowLine(154,56)(153,-22)
    \ArrowLine(348,58)(349,-13)
    \Text(153,251)[lb]{\Large{\Black{$\hat{1}$}}}
    \Text(345,251)[lb]{\Large{\Black{$\hat{2}$}}}
    \Text(248,144)[lb]{\Large{\Black{$\hat{P}$}}}
    \ArrowLine(386,81)(446,23)
    \Text(151,-45)[lb]{\Large{\Black{$5$}}}
    \Text(351,-45)[lb]{\Large{\Black{$4$}}}
    \Text(465,13)[lb]{\Large{\Black{$3$}}}
  \end{picture}
}
}
\end{center}
\label{5pt}
\caption{\it Recursive diagram for the five-point $\overline{\mathrm{MHV}}$ amplitude.}
\end{figure}
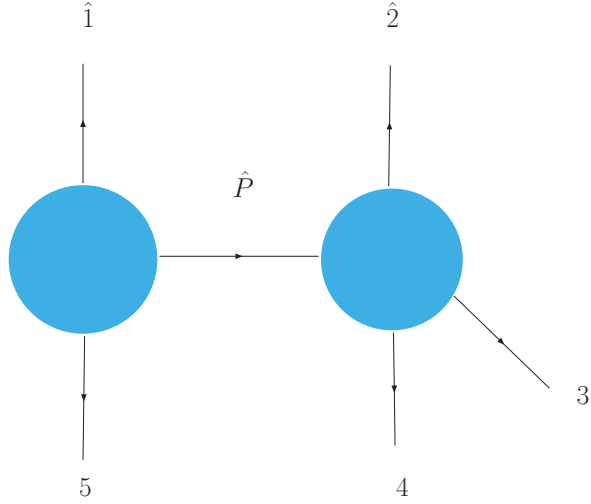

In this case, the two amplitudes are 
\beqa
\cA_L & = & 
{
\delta^{(4)} ( \hat{1} + 5 + \hat{P} ) \, 
\delta^{(8)} ( \hat{\eta}_1 \l_1 + \eta_5 \l_5 + \eta_{\hat{P}} \l_{\hat{P}}  ) 
\over 
\lan 1 \hat{P} \ran \lan \hat{P}  5 \ran \lan 5 1 \ran 
} 
\ , 
\\ \nonumber
\cA_R & = &  {\delta^{(4)} ( \hat{2} + 3 + 4  - \hat{P} ) \, 
\delta^{(8)} ( - \eta_{\hat{P}} \l_{\hat{P}}  + \eta_2 \hl_2 + \eta_3 \l_3 + \eta_4 \l_4 ) \over
\lan  \hat{P} \hat{2} \ran \lan \hat{2}  3 \ran \lan 3 4 \ran \lan 4 \hat{P}   \ran  
} 
\ .  
\eeqa
As usual, the product of two fermionic delta functions in $\cA_L$ and $\cA_R$ 
generates a delta function which imposes conservation of the supermomentum, 
$\delta^{(8)} ( \sum_{i} \eta_i \l_i)$.  

In order to simplify the expression of the amplitude  
it proves convenient to use the identity
\beq
\lan 34 \ran \lan\hat{P} 2 \ran \lan \hat{2} 3 \ran \lan 4 \hat{P} \ran  \ = \ 
{[ 4 \hat{P} ] [\hat{P} 2] [23] [3 4] \over [34]^4} \lan2 \hat{P}\ran^4 \ ,
\eeq
which is a consequence of momentum conservation.
One further notices that $\lan1 | \hat{P} | 4 ] = \lan15\ran [54]$, 
$\lan5 | \hat{P} | 2 ] = \lan51\ran [12]$ so that 
\beq
{1\over P_{15}^2} 
{1\over  \lan 1 \hat{P} \ran \lan \hat{P}  5 \ran \lan 5 1 \ran 
\lan  \hat{P} \hat{2} \ran \lan \hat{2}  3 \ran \lan 3 4 \ran \lan 4 \hat{P}   \ran  } \, = \, 
{1\over \prod_{i=1}^{5} [i i+1] } {[34]^4 \over    \lan 15 \ran^4 \lan \hat{2} \hat{P}\ran^4}
\ .  
\eeq
It is then easy to reproduce known component amplitudes from the recursive diagram in Figure 2. 
For practical evaluation purposes,  it is also convenient to use 
\beq
\delta^{(8)} \Big( \sum_i \eta_i^A \l_{i, \a} \Big) = 
{1\over 16}\prod_{A=1}^4 \sum_{i, j} \eta^A_i \eta^A_j \lan i j \ran
\ , 
\eeq
in order to extract the relevant contribution from the fermionic delta functions
\beq
\label{susu}
\int\!\!d^4\eta_{\hat{P}} \ \delta^{(8)} ( \hat{\eta}_1 \l_1 + \eta_5 \l_5 + \eta_{\hat{P}} \l_{\hat{P}}  ) \, 
\delta^{(8)} ( - \eta_{\hat{P}} \l_{\hat{P}}  + \eta_2 \hl_2 + \eta_3 \l_3 + \eta_4 \l_4 )
\ . 
\eeq
A few examples are in order.

For the split-helicity gluonic amplitude,  one picks from \eqref{susu}
the contribution proportional to $(\eta_1)^4 (\eta_2)^4 $, with the result 
\beq
\cA ( 5^-_{g} , 1^-_{g} , 2^-_{g} , 3^+_g , 4^+_{g} ) \ = \ i\,  {[34]^3\over 
[23][45][51][12]} 
\ . 
\eeq
For the gluonic amplitude with helicities $(5^- 1^- 2^+ 3^- 4^+)$  one picks from 
\eqref{susu} the coefficient of $(\eta_5)^4 (\eta_1)^4$. 
Further using that $ \lan 3 \hat{P} \ran^4 / \lan \hat{2} \hat{P} \ran^4 
= [24]^4  / [34]^4$, one quickly arrives at 
\beq
\cA ( 5^-_{g} , 1^-_{g} , 2^+_{g} , 3^-_g, 4^+_{g} ) \ = \ 
i\, {[24]^4\over 
[23][34] [45][51][12]} 
\ . 
\eeq
One could further proceed and consider amplitudes involving fermions and scalars. 
Consider for example the amplitude $ ( 5^-_{f} , 1^-_{g} , 2^-_{g} , 3^+_f, 4^+_{g} )$. 
Proceeding as before, the fermionic integrations produce a factor of 
$\lan 51 \ran^3 \lan\hat{P} 1 \ran \lan\hat{P} \hat{2} \ran^3 \lan \hat{2} 3 \ran$. 
Standard manipulations lead to $\lan 1 \hat{P} \ran  \lan \hat{2} \hat{P} \ran = 
\lan 1 5 \ran [52] / ( \lan 34 \ran [34] $, $\lan \hat{2} 3 \ran = - [45] \lan 34 \ran / [25] $, 
and one quickly finds that 
\beq
\cA ( 5^-_{f} , 1^-_{g} , 2^-_{g} , 3^+_f , 4^+_{g} ) \ = \ i\,  {[34]^3[45] \over 
[12] [23][34] [45][51]} 
\ , 
\eeq
in agreement with results of \cite{GK}. 

A further check is the derivation of  a four fermion amplitude
$ ( 5^-_{f_1} , 1^-_{f_2} , 2^+_{f_1} , 3^-_g, 4^+_{f_2} )$, where $f_1$ and $f_2$ denote fermions belonging to two different $\cN=1$ supermultiplets. 
Similar manipulations lead to the result 
\beq
\cA ( 5^-_{f_1} , 1^-_{f_2} , 2^+_{f_1} , 3^-_g, 4^+_{f_2} ) \ = \ 
i\,  {[45] [12] [24]^2  \over 
[12] [23] [34] [45][51]} 
\ , 
\eeq
in agreement with results of \cite{Luo:2005rx}.

\section{Proof of tree-level covariance}

In this section we wish to use a supersymmetric generalisation of the BCF recursion relations 
\cite{bcf,bcfw} to show that the 
tree-level $S$-matrix of $\mathcal{N}=4$ SYM
is covariant under dual superconformal transformations.
Here we will focus on the dual inversions of the dual superconformal group. 
As explained earlier,  it is most convenient to combine all amplitudes of a fixed total helicity 
and fixed number of external lines with the help of the dual superspace into one superamplitude, 
which is a natural generalisation of Nair's MHV superamplitude \eqref{NairMHV}. 
It is this superamplitude that we expect to transform uniformly, while the
component amplitudes usually do not have simple transformation properties under inversions
except for the split-helicity amplitudes \cite{dhks}.

Now assuming that all superamplitudes with up to $n$ external legs transform covariantly,  
we wish to use superspace generalisations of BCF recursion relations to show that all superamplitudes with $n+1$ legs also transform covariantly, and hence, by induction, 
that all superamplitudes with arbitrary numbers of external legs transform covariantly.
We will achieve this by showing in the following that actually each diagram in the recursion relation has
the correct covariant transformation behaviour, inherited from the transformation properties of the two subamplitudes entering the recursion diagram, the propagator,  and the bosonic and fermionic 
delta functions.

While the transformations of the region momenta $x_i$'s are  unique, there is a normalisation
ambiguity in the definition of inversions of the spinor variables $\lambda_i$'s.
In~\cite{dhks} the transformations of the spinors under a conformal
inversion were chosen to be $\lambda_{i}^\alpha \rightarrow
(x_i^{-1})_{\dot \alpha \beta} \lambda_i^\beta$. In the proof of
superconformal covariance of tree-level amplitudes constructed using
BCF recursion relations, it is however more useful to keep
the transformation of $\lambda_i$ more general and fix the normalisations later. 
We therefore consider the transformation
\begin{align}
\label{gentran2}
  \lambda_i^\alpha \rightarrow {x_{i}^{\dot \alpha \beta} \lambda_{i, \beta}
    \over \kappa_i}
    \ , 
\end{align}
and keep $\kappa_i$ arbitrary and local ({\it i.e.} they can have different
values, {\it e.g.} $x_i^2, x_{i+1}^2$ or $\sqrt{x_i^2x_{i+1}^2}$ for
different points $i$) although as we will see, we will be forced to
fix the factors $\kappa_i$ and $\kappa_{i+1}$ of the shifted momenta.
In order to complete the proof,  we consider the transformation
properties of amplitudes under this more general transformation.

By considering the explicit
expression for the tree-level MHV superamplitude ~(\ref{NairMHV}), one sees that it
transforms as
\begin{align}\label{MHVtrans2}
  \cA_{\rm MHV}(1,2,\dots,n)\  \rightarrow \ \cA_{\rm MHV}(1,2,\dots,n) \  \prod_{k=1}^n
  {\kappa_k^2    \over x_k^2} \ .
\end{align}
Now we wish to show recursively that in fact all 
tree-level superamplitudes transform in this way
under dual conformal inversions. 

Consider building a superamplitude recursively from two superamplitudes with
fewer legs, both of  which transform like the MHV amplitude above
under~(\ref{gentran2}) (see Figure 3), 
\begin{align}\label{altrans}
  \cA_L(j+1,j+2,\dots,\hat i,\hat P)&\rightarrow{\kappa_{j+1}^2 \dots
  \kappa_{i-1}^2 \hat \kappa_i^2 \hat \kappa_P^2 \over x_{j+1}^2 \dots
  x_i^2 \hat x_{i+1}^2}\cA_L(j+1,j+2,\dots,\hat i, \hat P)\ , \\\label{artrans}
 \cA_R(\widehat{i+1},i+2,\dots,j,-\hat P)&\rightarrow {\hat \kappa_{i+1}^2 \kappa_{i+2}^2 \dots
  \kappa_{j}^2 \hat \kappa_P^2 \over \hat x_{i+1}^2 x_{i+2}^2 \dots
  x_{j+1}^2 } \cA_R(\widehat{i+1},i+2,\dots,j,-\hat P)
  \ . 
\end{align}

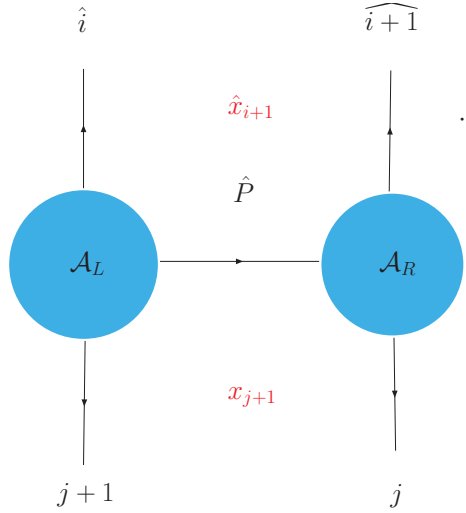
\begin{figure}
\begin{center}
\scalebox{0.60}{
\fcolorbox{white}{white}{
  \begin{picture}(458,316) (24,-64)
    \SetWidth{0.5}
    \SetColor{CornflowerBlue}
    \Vertex(153,89){46.69}
    \Vertex(347,89){44.55}
    \SetColor{Black}
    \ArrowLine(153,137)(153,212)
    \ArrowLine(345,135)(346,211)
    \ArrowLine(201,91)(301,91)
    \ArrowLine(154,41)(153,-37)
    \ArrowLine(348,43)(349,-28)
    \Text(248,129)[lb]{\Large{\Black{$\hat{P}$}}}
    \Text(150,236)[lb]{\Large{\Black{$\hat{i}$}}}
    \Text(331,236)[lb]{\Large{\Black{$\widehat{i+1}$}}}%
    \Text(347,-64)[lb]{\Large{\Black{$j$}}}
    \Text(137,-63)[lb]{\Large{\Black{$j+1$}}}
    \Text(390,177)[lb]{\Large{\Black{$\cdot$}}}
    \Text(426,155)[lb]{\Large{\Black{$\cdot$}}}
    \Text(449,114)[lb]{\Large{\Black{$\cdot$}}}
    \Text(452,66)[lb]{\Large{\Black{$\cdot$}}}
    \Text(440,25)[lb]{\Large{\Black{$\cdot$}}}
    \Text(407,-8)[lb]{\Large{\Black{$\cdot$}}}
    \Text(77,177)[lb]{\Large{\Black{$\cdot$}}}
    \Text(40,150)[lb]{\Large{\Black{$\cdot$}}}
    \Text(24,114)[lb]{\Large{\Black{$\cdot$}}}
    \Text(26,66)[lb]{\Large{\Black{$\cdot$}}}
    \Text(39,26)[lb]{\Large{\Black{$\cdot$}}}
    \Text(72,-8)[lb]{\Large{\Black{$\cdot$}}}
    \Text(145,83)[lb]{\Large{\Black{$\mathcal{A}_L$}}}
    \Text(340,83)[lb]{\Large{\Black{$\mathcal{A}_R$}}}
     \Text(244.65,180.61)[lb]{\Large{\Red{$\hat{x}_{i+1}$}}}
    \Text(244.65,-0.61)[lb]{\Large{\Red{$x_{j+1}$}}}
  \end{picture}
}
}
\end{center}
\caption{\it Generic recursion diagram used in the proof of covariance.}
\end{figure}

In the recursion we will make use of the shift denoted by $[i\, i+1 \ran$, {\it i.e.}
\begin{align}
{\hat {\tilde \lambda}}_i = \tilde \lambda_i +z \tilde \lambda_{i+1} \ , \qquad \qquad
\hat \lambda_{i+1}= \lambda_{i+1} -z  \lambda_i
\ , 
\end{align}
with all other spinors unchanged.

A couple of comments are in order before we proceed. First of all, 
consider the spinor variables $\lambda_{\hat{P}}$ and $\tilde{\lambda}_{\hat{P}}$ 
of the internal on-shell leg $\hat{P}$.
If we use the DHKS transformation of $\lambda_i$ and do not introduce $\kappa_i$,
then from the point of view of $\mathcal{A}_L$ the spinor $\lambda_{\hat{P}}$ would transform under inversions into
$\hat{x}_{i+1} \lambda_{\hat{P}}/(\hat{x}_{i+1})^2$, 
and from the point of view $\mathcal{A}_R$ into 
$x_{j+1} \lambda_{\hat{P}}/(x_{j+1})^{2}$, which are not compatible. 
This is why we have introduced an arbitrary factor into the
$\lambda$ transformations. For the spinor $\lambda_{\hat{P}}$ we have
 \begin{align}\label{gentran}
   \lambda_{\hat{P}}^\alpha \rightarrow {\hat{x}_{i+1}^{\dot \alpha \beta} \lambda_{\hat{P}, \beta}
     \over \hat{\kappa}_P} \, = \, 
     {x_{j+1}^{\dot \alpha \beta} \lambda_{\hat{P}, \beta}
     \over \hat{\kappa}_P} 
     \  .
 \end{align}

Secondly, the two superamplitudes
$\mathcal{A}_L$ and $\mathcal{A}_R$ above depend on unshifted momenta but also on the shifted momenta $\hat{p}_i$, $\hat{p}_{i+1}$ and $\hat{P}$. By assumption 
these amplitudes are covariant under inversions of the corresponding sets of shifted and
unshifted momenta using the assignments of  region momenta in Figure 3. 
On the other hand, every recursive diagram depends only on unhatted quantities due to the fact
that hatted quantities depend via $z$ only on unhatted quantities. To be more specific,
$z$ for the recursive diagram given above has to be set to the solution of the equation
\be
(P - z \lambda_{i} \tilde{\lambda}_{i+1})^2=0
\ , 
\ee
which is $z_P= P^2/[ i+1 | P | i \rangle$,  where $P=P_R:=\sum_{l=i+1}^{j} p_l$.
It can easily be checked that the two seemingly different definitions of the transformations of hatted quantities as defined above and
as {\it inherited} from the unhatted quantities, combined with the  appropriate transformation of
$z= z_P$, are actually identical. For the purpose of the proof it is more convenient to work
with the inversions of hatted quantities as defined above, hence we will use those in what follows, 
but the reader should keep in mind that this is completely equivalent to performing all transformations on unhatted quantities.

An important fact to note at this point is that, whereas so far we
have kept 
the $\kappa_i$ arbitrary, the $[i \, i+1\rangle$ shift in fact fixes the
transformation under inversions of $\lambda_i $ and $\lambda_{i+1}$.
To see this,  note that $\hat \lambda_i = \lambda_i$ and so the
transformation $\hat \lambda_i^\alpha \rightarrow {x_{i}^{\dot \alpha
    \beta} \hat \lambda_{i, \beta}
   / \hat \kappa_i}$ must be consistent with $ \lambda_i^\alpha
  \rightarrow {x_{i}^{\dot \alpha 
    \beta} \lambda_{i, \beta}
    / \kappa_i}$  under inversions, 
requiring $\hat \kappa_i=\kappa_i$.  
  A more complicated consistency condition comes from considering the
  transformation of $\hat \lambda_{i+1}^\alpha$
 and comparing with the transformation of
  $\lambda_{i+1} -z\lambda_{i}$. Here the factors $\kappa$ will in general be
  functions of the region momenta $x$ and so the  shifted factors
  $\hat \kappa$ are simply
  the same function of the shifted region momenta $\hat x$.
One solution of these conditions is 
\begin{align}\label{ki}
  \kappa_i=x_i^2 \qquad \qquad \kappa_{i+1}=x_{i+1}^2\ ,\qquad
  \Rightarrow \qquad   \hat \kappa_i=x_i^2 \qquad \qquad \hat \kappa_{i+1}=\hat x_{i+1}^2
  \ , 
\end{align}
   which we assume from now on.

Now, in order to use an induction proof on the number of legs, we
consider the contribution to the superamplitude given by the 
recursive  diagram in Figure 3, 
\begin{align}
\label{sn}
  \int\!{d^4 P \over P^2}  \int\!d^4 \eta_{\hat{P}} \,
  &\delta^{(4)}(P_L+P)\,\delta^{(8)}(\hat{\Lambda}_L +\lambda_{\hat{P}} 
    \eta_{\hat{P}}) 
    \\ \nonumber
&\!\!\!\!\!\!\!\!\times\delta^{(4)}(P_R-P)\,\delta^{(8)}(\hat{\Lambda}_R -\lambda_{\hat{P}} 
  \eta_{\hat{P}}) \, A_L \, A_R\\ \nonumber
=\ \delta^{(4)}(P_L+P_R)&\delta^{(8)}(\Lambda_L+\Lambda_R){1 \over P^2}
\delta^{(4)}(\lan \lambda_{\hat{P}} \hat \Lambda^A_L \ran ) \, A_L \, A_R\ , 
\end{align}
where we have defined amplitudes with momentum conservation and 
supermomentum conservation  delta functions removed as $A_{L,R}$, 
\begin{align}
\cA \ = \ \delta^{(4)}\Big(\sum_k p_k\Big) \, \delta^{(8)}\Big(\sum_k \eta_k \lambda_k\Big) \, A\ .
\end{align}
We have also introduced the shorthand notation 
   $\Lambda_L := \sum_{ l=j+1}^{i} \eta_l \lambda_l$,  $\hat{\Lambda}_L := 
\sum_{ l=j+1}^{\hat{i}} \eta_l \lambda_l$, and $P_L :=  \sum_{ l=j+1}^{i} \lambda_l \lt_l $ as usual.
Similarly,  we have defined  $\Lambda_R := \sum_{ l=i+1}^{j} \eta_l \lambda_l = - \Lambda_L$,  
$\hat{\Lambda}_R :=  \sum_{ l=\widehat{i+1}}^{j} \eta_l \lambda_l = - \hat{\Lambda}_L$, and 
$P_R :=  \sum_{ l=i+1}^{j} \lambda_l \lt_l = - P_L$.
Notice  also that $\hat{\Lambda}_L = \hat{\theta}_{i+1} - \theta_{j+1}$. 
Finally, we observe that in the last line of \eqref{sn}, $\eta_{\hat{P}}$ appearing inside 
$A_L$ and $A_R$ should be thought of as the solution of the equation 
$\hat{\Lambda}_L +\lambda_{\hat{P}} 
    \eta_{\hat{P}}=0$. 

Using~(\ref{gentran}) and  the standard transformations \eqref{xtran} and \eqref{thtran} 
of the $x_i$ and the $\theta_i$  under inversions,  we find
\begin{align}
  {1\over P^2} = {1 \over (x_{i+1}-x_{j+1})^2}&\rightarrow x_{i+1}^2 x_{j+1}^2 {1\over P^2}
  \ , 
  \\
\delta^{(4)}(\lan \lambda_{\hat{P}} \hat{\Lambda}^A_L \ran ) &\rightarrow
{1 \over \hat \kappa_P^4}   \delta^{(4)}(\lan \lambda_{\hat{P}} \hat \Lambda^A_L \ran ) 
\ , 
\end{align}
and, hence, together with~(\ref{altrans}) and (\ref{artrans}) we infer  
that the recursive diagram in Figure 3 transforms with weight
\begin{align}\label{genweight}
&x_{i+1}^2 x_{j+1}^2 \ {1\over \hat \kappa_P^4} \  {\kappa_{j+1}^2 \dots
  \kappa_{i-1}^2 \hat \kappa_i^2 \hat \kappa_P^2 \over x_{j+1}^2 \dots
  x_i^2 \hat x_{i+1}^2}\  {\hat \kappa_{i+1}^2 \kappa_{i+2}^2 \dots
  \kappa_{j}^2 \hat \kappa_P^2 \over \hat x_{i+1}^2 x_{i+2}^2 \dots
  x_{j+1}^2 } \nonumber \\
=  \ &\prod_{k=1}^n {\kappa_k^2 \over x_k^2} \  {\hat \kappa_i^2 \hat
    \kappa_{i+1}^2 \over \kappa_i^2 \kappa_{i+1}^2} \ 
  {(x_{i+1})^4 \over (\hat x_{i+1})^4} \nonumber \\
=\ &\prod_{k=1}^n {\kappa_k^2 \over x_k^2}
\ , 
\end{align}
as required. The last equality follows directly from the values of
$\kappa_i,\ \kappa_{i+1}$ and $\hat \kappa_{i+1}$ given in~(\ref{ki}).

In the analysis of the covariance properties of a generic
tree amplitude using recursion relations, we may encounter  
diagrams where either $\cA_L$ or $\cA_R$ is the
three-point anti-MHV amplitude given in~\eqref{3ptmhvbar}.  This class of  diagrams is  somewhat special 
since \eqref{3ptmhvbar} does not contain the standard supermomentum conservation delta function. 
However, we have shown in  \eqref{rj} that \eqref{3ptmhvbar} transforms in the correct way under dual superconformal symmetry, hence recursive diagrams involving a three-point anti-MHV amplitude are in fact  
not  special from the point of view of the covariance properties. 
For completeness, we discuss now how a generic diagram in this class transforms 
under conformal inversions.  

Let  $\cA_R$ then be the three-point anti-MHV
amplitude. Then the generic recursive diagram in this class is of the form 
\begin{align}
\label{speciall}
 & \int {d^4 P \over P^2}  \int d^4 \eta_{\hat{P}} \,
  \delta^{(4)}(P_L+P)\,\delta^{(8)}(\hat{\Lambda}_L +\lambda_{\hat{P}} 
    \eta_{\hat{P}})\,  \cA_L \nonumber
\\
&\hspace{60pt}\times\delta^{(4)}(P_R-P)\,{\delta^{(4)}( \eta_{\hat{P}} [\widehat{i+1} j] \, + \, \eta_{i+1}
  [ j \, -  \! \hat P] \, + \, \eta_j [-\hat P \, \widehat{i+1}] ) \over [\widehat{i+1} \,j]\,
  [ j -\hat P] \, [-\hat P \,\widehat{i+1}]}  \nonumber  \\[10pt]
= \ &\delta^{(4)}(P_L+P_R)\delta^{(8)}(\L_L+\L_R)\, 
{1 \over P^2}\, 
\, \cA_L \,  { [\widehat{i+1} j]^3 \over  
  [ j -\hat P] \, [-\hat P \,\widehat{i+1}]}\ ,
\end{align}
where $j=i+2$ since we are dealing with a three-point amplitude on the
right.  Now the conjugate spinors transform as 
\begin{equation}
\tilde \lambda_{k, {\dot{\a}}} 
\rightarrow -{\kappa_k \over x_k^2 x_{k+1}^2} x_{k,\dot\beta \alpha}
\tilde \lambda^{\dot \beta}_k
\end{equation}
 under inversions (for consistency with
the transformation of $p_k=\lambda_k\tilde\lambda_k$),  
hence  the square
brackets transform as 
\begin{align}
  [k\,k+1] \rightarrow {\kappa_k \kappa_{k+1}  \over x_{k}^2 \,
    x_{k+1}^2 \,x_{k+2}^2} [k\,k+1]     \ .
\end{align}
We then find that the diagram transforms with weight
  \begin{align}
&x_{i+1}^2 x_{j+1}^2  \  {\kappa_{j+1}^2 \dots
  \kappa_{i-1}^2 \hat \kappa_i^2 \hat \kappa_P^2 \over x_{j+1}^2 \dots
  x_i^2 \hat x_{i+1}^2}\      {\hat \kappa_{i+1}^2 
  \kappa_{j}^2  \over \hat \kappa_P^2 \hat x_{i+1}^2 x_{j}^2 
  x_{j+1}^2 } \nonumber \\
=\ &\prod_{k=1}^n {\kappa_k^2 \over x_k^2}
\end{align}
(using~(\ref{ki})),  precisely as required. 

In conclusion, we have found that each recursive diagram with shifts $[i \, i+1 \rangle$ 
contributing to  a generic superamplitude transforms covariantly under dual conformal inversions 
once we assume that  
$\cA_L$ and $\cA_R$ transform as superamplitudes.  
From this, and from the arbitrariness  of the choice of the legs $i$ and $i+1$, 
we conclude by induction  that all tree-level superamplitudes
in $\mathcal{N}=4$ SYM transform covariantly as the MHV
amplitudes, {\it i.e.} as in \eqref{MHVtrans2}. 

Note that in the conventions of \cite{dhks}
we would have to set $\kappa_k = x_k^2$ for all $k$,  
and the last line of (\ref{genweight}) would become just
\be
\prod_{k=1}^n x_k^2 \, ,
\ee
which shows that this recursive diagram and hence the whole amplitude transforms uniformly
with weight one under inversions.

\section{Covariance of the coefficients  of one-loop amplitudes}

In this section we discuss how generic one-loop amplitudes in $\mathcal{N}=4$ SYM inherit 
the transformation properties under dual superconformal symmetry  from the tree-level amplitudes. 
It is a well known fact that all one-loop amplitudes in $\mathcal{N}=4$ SYM can be
expanded in a basis of integral functions which consists only of so-called one-loop scalar
boxes,  with coefficients that are rational functions of the kinematic variables \cite{bddk}.
We will show in the following that the coefficients of the  expansion%
\footnote{The precise definition of the basis  will be given shortly.}
of an arbitrary  $\cN=4$ SYM superamplitude in terms of box functions 
are given by conformally covariant functions  which transform in  the same way as the corresponding tree-level superamplitude.   

This claim is motivated by the special form of the coefficients in the expansion of 
the split-helicity gluonic amplitudes at one loop calculated in \cite{vittorio} and 
\cite{dissolving}. Inspection of the results of these papers shows that these coefficients 
are covariant under conformal inversions, as they are made of spinor brackets consisting of strings of spinors always belonging to adjacent legs.  
Another simple example is provided by the infinite sequence of one-loop MHV superamplitudes 
in $\cN=4$ SYM.  
This superamplitude was calculated in \cite{bddk}, and re-derived in \cite{bst} using 
$\cN=4$ supersymmetric MHV diagrams, and is written 
as a sum of  two-mass easy box functions, all with the same coefficient.%
\footnote{An explanation of why these box functions appear all with the same coefficient  
 -- equal to one, if one factors out the tree amplitude --  
was given in terms of the Wilson loop/MHV amplitudes duality in \cite{bht}.}
This coefficient is  equal to the tree-level MHV superamplitude, which is of course covariant.

Before we proceed,  it is important to make a comment on the basis of integral functions
that we expand in.
The natural basis to consider in the context of dual conformal symmetry 
is that given by the so-called scalar box functions
$F_i$,  which are (pseudo-)conformally invariant \cite{magic}, and are related to the more 
standard scalar box integrals $I_i$ by a kinematic prefactor \cite{bddk}.
The external momenta at the four corners of a given box function, $K_1, K_2, K_3$ and $K_4$ (see Figure 4),  are in general sums of momenta $p_i$ of external
particles of the $n$-point amplitude under consideration. Alternatively, the momenta 
$K_{1\ldots4}$ can be expressed in terms of the region momenta 
$x_{1\ldots4}$ as in Figure 4, {\it e.g.} $K_1 = x_{12}$, where $x_{ij} := x_i-x_j$.
Then, up to a numerical constant, the relation between the $F$'s and the $I$'s  is%
\footnote{In \eqref{bpm}  we use a collective index $i$ to denote the box function with external momenta 
$K_{1\ldots 4}$,  as in  Figure 4.}
\beqa
\label{bpm}
I_i &= &\frac{F_i}{\sqrt{R_i}} \ , \nonumber \\
R_i&=& (x_{13}^2 x_{24}^2)^2 - 2 x_{13}^2 x_{24}^2  x_{12}^2  x_{34}^2  
- 2 x_{13}^2 x_{24}^2  x_{23}^2  x_{41}^2 +(x_{12}^2 x_{34}^2 - x_{23}^2 x_{41}^2)^2 \ .
\eeqa
It will be useful for later to quote here the transformation of the kinematic factor 
$\sqrt{R_i}$ under dual conformal  inversions:
\be\label{rtrafo}
\sqrt{R_i} \ \to \  \frac{\sqrt{R_i}}{x_1^2 x_2^2 x_3^2 x_4^2} \, .
\ee
Obviously we can expand the amplitude in either basis. We write (schematically),   
\be
\mathcal{A}_{\mathrm{1-loop}} \ =  \ \sum \mathcal{B}_i\,  I_i \ = \ 
\sum \tilde{\mathcal{B}}_i \, F_i \, .
\ee
We will show that it is the supersymmetric generalisation of the coefficients 
$\tilde{\mathcal{B}}_i = \mathcal{B}_i/\sqrt{R_i}$ that have uniform covariant transformation properties under
dual superconformal transformations just as the corresponding tree-level amplitudes, while
the $\mathcal{B}_i$ have mixed transformation properties.

\begin{figure}

\begin{center}
\scalebox{0.60}{
\fcolorbox{white}{white}{
    \begin{picture}(387,378) (110,-63)
    \SetWidth{0.5}
    \SetColor{Black}
    \ArrowLine(227,214)(370,214)
    \ArrowLine(369,52)(233,52)
    \Text(293,133)[lb]{\Large{\Black{$x_5,\ \theta_5$}}}
    \COval(211,57)(25,25)(0){Black}{CornflowerBlue}
    \COval(390,59)(25,25)(0){Black}{CornflowerBlue}
    \COval(385,216)(25,25)(0){Black}{CornflowerBlue}
    \COval(212,217)(25,25)(0){Black}{CornflowerBlue}
    \ArrowLine(209,83)(210,192)
    \ArrowLine(388,191)(388,84)
    \Text(448,139)[lb]{\Large{\Red{$x_2,\ \theta_2$}}}
    \Text(152,139)[lb]{\Large{\Red{$x_4,\ \theta_4$}}}
    \Text(293,260)[lb]{\Large{\Red{$x_1,\ \theta_1$}}}
    \Text(293,-5)[lb]{\Large{\Red{$x_3,\ \theta_3$}}}
    \ArrowLine(193,235)(153,277)
    \ArrowLine(404,232)(447,277)
    \ArrowLine(404,37)(447,-13)
    \ArrowLine(194,36)(153,-13)
    \Text(467,298)[lb]{\Large{\Black{$K_1$}}}
    \Text(462,-63)[lb]{\Large{\Black{$K_2$}}}
    \Text(111,-62)[lb]{\Large{\Black{$K_3$}}}
    \Text(110,299)[lb]{\Large{\Black{$K_4$}}}
    \Text(291,221)[lb]{\Large{\Black{$l_1$}}}
    \Text(394,138)[lb]{\Large{\Black{$l_2$}}}
    \Text(216,137)[lb]{\Large{\Black{$l_4$}}}
    \Text(300,34)[lb]{\Large{\Black{$l_3$}}}
    \Text(201,213)[lb]{\Large{\Black{$\cA_1$}}}
    \Text(377,213)[lb]{\Large{\Black{$\cA_2$}}}
    \Text(377,52)[lb]{\Large{\Black{$\cA_3$}}}
    \Text(201,52)[lb]{\Large{\Black{$\cA_4$}}}
  \end{picture}
}
}
\end{center}

\caption{\it Quadruple cut of a one-loop superamplitude in $\mathcal{N}=4$ SYM. The four blobs represent 
tree-level $\mathcal{N}=4$ superamplitudes. The $K_{1\ldots4}$ correspond to sums of
momenta $p_i$ of the external particles.}
\end{figure}
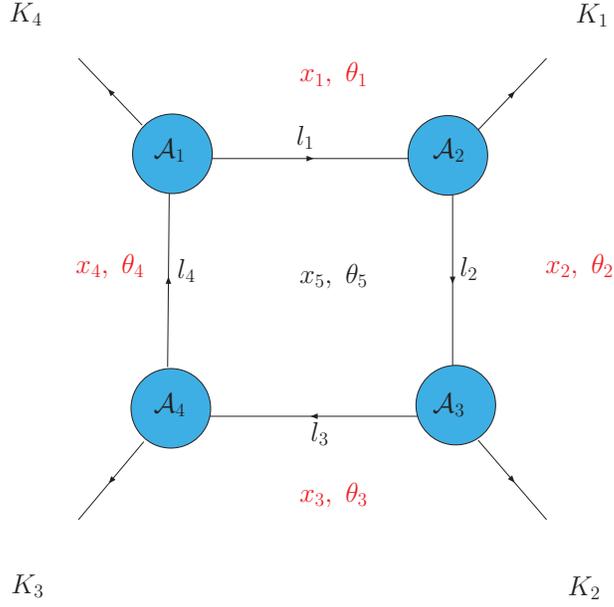

In order to prove this statement,   we now  discuss in more detail  quadruple cuts of one-loop amplitudes.  
As mentioned above, all one-loop amplitudes in $\cN=4$ SYM are expressed in terms of box 
functions only \cite{bddk} and their coefficients can be calculated most efficiently with 
quadruple cuts \cite{bcfgen}. This technique allows one to calculate 
the coefficients of the box functions one by one, and  the problem of finding general one-loop amplitudes in $\mathcal{N}=4$ SYM is reduced to a purely algebraic one,
as the coefficients turn out to be given by products of four tree-level amplitudes.
Importantly, quadruple cuts freeze the one-loop integration completely and, hence, one 
can stay in four dimensions, without introducing any regularisation. 

A generic quadruple cut box is of the form
\beq
\int\! d^4 l \ \delta^{(+)} (l^2)  \delta^{(+)} ( (l- K_1)^2)\delta^{(+)} ((l- K_1 -  K_2)^2)
\delta^{(+)} ( (l + K_4)^2)
\ , 
\eeq
or, re-expressing it in terms of the region momenta in Figure 4, 
\beq
\int\! d^4 x_5 \   \delta^{(+)} ( x_{51}^2)\delta^{(+)} (x_{52}^2)
\delta^{(+)} ( x_{53}^2) \delta^{(+)} (x_{54}^2)
\ .  
\eeq
Under conformal inversions, the delta functions transform 
in  the same way as ordinary propagators, 
except for the sign of the energy component, which is flipped, so that 
$\delta^{(+)} \big ( ( x - y)^2 \big) \to x^2 y^2 \, \delta^{(-)}  \big ( ( x - y)^2 \big)$. Therefore, 
\beqa
&&\int\!d^4 x_5 \   \delta^{(+)} ( x_{51}^2)\delta^{(+)} (x_{52}^2)
\delta^{(+)} ( x_{53}^2) \delta^{(+)} (x_{54}^2)   
\nonumber \\
& \to  &(x_1^2 x_2^2 x_3^2 x_4^2)  \int\!d^4 x_5 \   \delta^{(-)} ( x_{51}^2)\delta^{(-)} (x_{52}^2)
\delta^{(-)} ( x_{53}^2) \delta^{(-)} (x_{54}^2)
\ . 
\eeqa
Furthermore, the quadruple cut of the corresponding scalar box function $F$ is invariant
under dual conformal inversions.

The coefficient of the scalar box integral $I$  appearing in the expansion of the amplitude is 
then evaluated as \cite{bcfgen}
\beq
\label{coeff}
\cB  \ = \ {1 \over n_S} \sum_{S, J} n_J \, (\cA_1\cA_2 \cA_3 \cA_4)
\ , 
\eeq
where $n_S$ is the number of solutions $S$ to the cut condition, and the sum is extended to  particles of all spin $J$ in the $\cN=4$ theory  which can run in the loop.  $n_J$ is the number of particles of spin $J$. 
$\cA_i$, $i=1, \ldots, 4$ are the four tree-level amplitudes at the four corners of the quadruple cut, 
as in Figure 4.

In order to show in full generality that the coefficients of the one-loop superamplitudes
in $\cN=4$ SYM are dual superconformal covariant,
we have to generalise \eqref{coeff} in a supersymmetric way 
by lifting the amplitudes to superamplitudes, and introducing the appropriate fermionic 
delta functions which impose  supermomentum conservation at the four corners of
the diagram in Figure 4. This procedure will also lift the coefficient $\mathcal{B}$ in
\eqref{coeff} to an appropriate supercoefficient. Doing this, 
we get the following expression for the quadruple cut,%
\footnote{An equivalent  supersymmetric extension of the quadruple cuts 
has been introduced in  \cite{dhksgen}. There it was used to calculate explicitly supercoefficients of 
NMHV one-loop amplitudes and four-mass box coefficients of NNMHV one-loop amplitudes, and, 
furthermore, it was checked 
that these supercoefficients are covariant under dual superconformal transformations.}
which implicitly defines the supercoefficient $\cB$:
\begin{align}
\label{pre}
&\delta^{(4)}\big(\sum_{i=1}^{4} K_i \big) \delta^{(8)}\big(\sum_{i=1}^{4} \Lambda_i\big) \ \cB \\ 
:= \ 
&\delta^{(4)}\big(\sum_{i=1}^{4} K_i\big) 
{1\over n_S} \sum_S 
\ 
\int\!\!\prod_{i=1}^{4}d^4\eta_{l_i} \ 
\delta^{(8)}(\lambda_{l_2}\eta_{l_2}-\lambda_{l_1} \eta_{l_1} +\theta_{12})
\delta^{(8)}(\lambda_{l_3}\eta_{l_3}-\lambda_{l_2} \eta_{l_2} +\theta_{23})\nonumber\\[0pt]
& \ \ \ \ \ \ \ \ \ \ \ \ \ \ \ \ \ \ \ \ \ \ \ \
\delta^{(8)}(\lambda_{l_4}\eta_{l_4}-\lambda_{l_3} \eta_{l_3} +\theta_{34})
\delta^{(8)}(\lambda_{l_1}\eta_{l_1}-\lambda_{l_4} \eta_{l_4} +\theta_{41}) \ A_1 A_2
A_3 A_4   \ , 
 \nonumber
\end{align}
where, as previously, $A_i$ are the relevant superamplitudes with the
momentum and supermomentum delta functions removed.%
\footnote{In \eqref{pre} we consider the case where each of the four tree superamplitudes provides an eight-dimensional delta function of supermomentum conservation. 
The case where some of the  tree amplitudes 
are three-point  $\overline{\rm MHV}$ superamplitudes requires a special treatment, similar to that
presented in \eqref{speciall} in the proof of covariance of the tree-level recursion relation. }
The cut loop momenta are defined as $l_i := \l_{l_i} \lt_{l_i}$, $i =  1, \ldots, 4$, and we set 
$\theta_{ij} := \theta_i - \theta_j$. We have also defined $\Lambda_i := \sum_{i \in K_i} \eta_i \lambda_i$.  

Next,  we replace one of the fermionic delta functions with an overall
supermomentum conservation delta function, 
and then perform the $\eta_{l_1},\eta_{l_4}$ integrations to get
\begin{align}\label{oneloopa}
&\delta^{(4)}\big(\sum_{i=1}^{4} K_i \big) \delta^{(8)}\big(\sum_{i=1}^{4} \Lambda_i\big)
\\
&\times 
{1\over n_S} \sum_S 
\ 
\int\!\!\prod_{i=1}^{4}d^4\eta_{l_i} \ 
\delta^{(8)}(\lambda_{l_2}\eta_{l_2}-\lambda_{l_1} \eta_{l_1} +\theta_{12})
\delta^{(8)}(\lambda_{l_3}\eta_{l_3}-\lambda_{l_2} \eta_{l_2} +\theta_{23})\nonumber\\[0pt]
& \ \ \ \ \ \ \ \ \ \ \ \ \ \
\delta^{(8)}(\lambda_{l_4}\eta_{l_4}-\lambda_{l_3} \eta_{l_3} +\theta_{34}) \ A_1 A_2
A_3 A_4 \nonumber   \\[10pt]
=\ & 
\delta^{(4)}\big(\sum_{i=1}^{4} K_i \big) \delta^{(8)}\big(\sum_{i=1}^{4} \Lambda_i\big) 
\nonumber\\
&\times 
{1\over n_S} \sum_S 
\int\!\!d^4\eta_{l_2}d^4\eta_{l_3} \
\delta^{(4)}\big(\langle l_1 \theta_{15}^A \rangle \big)\ 
\delta^{(4)}\big(\langle l_4 \theta_{45}^{A} \rangle \big)\
\delta^{(8)}(\lambda_{l_3}\eta_{l_3}-\lambda_{l_2} \eta_{l_2} +\theta_{23}) \ A_1 A_2
A_3 A_4  \nonumber \\[10pt]
=\ &\delta^{(4)}\big(\sum_{i=1}^{4} K_i \big) \delta^{(8)}\big(\sum_{i=1}^{4} \Lambda_i\big)  
\ {1\over n_S} \sum_S
\delta^{(4)}\big(\langle l_1 \theta_{15}^A \rangle \big)\ 
\delta^{(4)}\big(\langle l_4 \theta_{45}^{A} \rangle \big)\
\langle l_2 l_3 \rangle^4\ A_1 A_2
A_3 A_4 \ .
\nonumber
\end{align}
We now consider the transformation of this expression under inversions. As in the
 proof of tree-level covariance presented earlier,  
 we make use of the more general form of the
transformations involving unspecified parameters $\kappa_i$ (see \eqref{gentran2}).%
\footnote{Note however that no  hatted  quantities appear here,  unlike the case of the recursion relation in Section 4.}

Under dual conformal inversions, the various quantities 
in (\ref{oneloopa}) transform as
\beqa
\label{trr}
&&  \delta^{(4)}\big(\langle l_1 \theta_{15}^A \rangle \big) \rightarrow
\Big({1 \over \kappa_{l_1}}\Big)^4 \delta^{(4)}\big(\langle l_1 \theta_{15}^A \rangle
  \big) \,\, ,\,\,
\delta^{(4)}\big(\langle l_4 \theta_{45}^{A} \rangle \big) \rightarrow
\Big({1 \over \kappa_{l_4}}\Big)^4 \delta^{(4)}\big(\langle l_4
\theta_{45}^{A} \rangle \big) \,\, ,  \nonumber \\
&& \langle l_2 l_3 \rangle \rightarrow  \Big( {x_5^2 \over \kappa_{l_2}
  \kappa_{l_3}}\Big)  \langle l_2 l_3 \rangle \,\, , \nonumber  \\
&& A_1 \rightarrow {\kappa_4^2 \kappa_{l_1}^2 \kappa_{l_4}^2 \over x_1^2 x_4^2
  x_5^2} \, A_1 \,\, , \,\, \qquad 
A_2 \rightarrow {\kappa_1^2 \kappa_{l_1}^2 \kappa_{l_2}^2 \over x_1^2 x_2^2
  x_5^2} \, A_2 \,\, , \nonumber \\
&& A_3 \rightarrow {\kappa_2^2 \kappa_{l_2}^2 \kappa_{l_3}^2 \over x_2^2 x_3^2
  x_5^2} \, A_3  \,\, , \,\, \qquad
A_4 \rightarrow {\kappa_3^2 \kappa_{l_3}^2 \kappa_{l_4}^2 \over x_3^2 x_4^2
  x_5^2} \, A_4 \, .
\eeqa
For the sake of brevity, in writing  the transformations of  $A_{1\ldots 4}$ we have included 
only the dependence on the transformation of the region momenta 
$x_{1\ldots5}$
because all other region momenta are just  spectators in this diagram -- 
any transformation properties with respect to them  are directly inherited 
from the superamplitudes entering the quadruple cut.

Inserting the transformations \eqref{trr} into (\ref{oneloopa}),  we see that the corresponding
(super)coefficient $\mathcal{B}$ transforms as
\begin{align}
\label{ra}
\mathcal{B} \ \to \ \mathcal{B} \,  {\kappa_1^2 \kappa_2^2 \kappa_3^2 \kappa_4^2 \over x_1^4 x_2^4 x_3^4
    x_4^4 }\ .
\end{align}
For any of the standard choices of the $\kappa$'s, the ratio in 
\eqref{ra}  would give 1,  and the coefficient $\cB$ would then be
invariant with respect to the transformation of the region momenta $x_{1\ldots4}$.

The $\mathcal{B}_i$'s  are the coefficients relevant for the expansion
in the scalar box integrals $I_i$ basis, which the quadruple cut actually calculates. 
As mentioned earlier,   from the point of view of dual conformal symmetry 
it is more natural to consider the transformation properties of the 
coefficients   $\tilde{\mathcal{B}} = \mathcal{B}/\sqrt{R}$ of the expansion in terms of 
scalar box functions $F_i$. 
The transformation of these  coefficients  is immediately  obtained using \eqref{ra} and
\eqref{rtrafo}, 
\begin{align}
\label{endd}
\tilde{\mathcal{B}}\  \to\  \tilde{\mathcal{B}} \,  {\kappa_1^2 \kappa_2^2 \kappa_3^2 \kappa_4^2 \over x_1^2 x_2^2 x_3^2 x_4^2}\ ,
\end{align}
which,  upon making the standard choice for the $\kappa_i$,  becomes
$\tilde{\mathcal{B}} \to  \tilde{\mathcal{B}}  \, x_1^2 x_2^2 x_3^2 x_4^2
$ . 
Reinstating the transformation properties of the spectator region momenta, \eqref{endd}  
shows that the supercoefficients $\tilde{\mathcal{B}}$ of the expansion of the superamplitude in terms of  the scalar box functions $F$'s transform covariantly under dual inversions just as
the tree-level superamplitudes, {\it i.e.} 
\beq
\tilde\cB \ \to \ \tilde\cB \, \prod_{i=1}^{n} {\kappa_i^2 \over x_i^2 }  
\ . 
\eeq

We would like to conclude with a few comments. 

{\bf 1.} 
By performing four-dimensional quadruple cuts we have by-passed 
the problem  of dimensionally regularising the theory (thus breaking conformal invariance). 
It is only when the cut box is lifted to a full, $D$-dimensional integral box function 
that infrared divergences appear  (and therefore need to be regulated). However, 
for the sake of determining the transformation properties of the 
coefficients of the box functions, one can remain in four dimensions. 
The MHV anomaly of \cite{dhks} is of course hiding inside the anomalous transformation properties under 
dual conformal transformations of the $D$-dimensional box functions.

{\bf 2. } 
It is amusing to note that the covariance of the integral coefficients of the one-loop amplitudes provides also an alternative proof that all tree-level superamplitudes with arbitrary total helicity 
are dual superconformal covariant. 
This is a simple consequence of the universal structure of infrared divergences of 
one-loop amplitudes
\begin{equation}
\mathcal{A}_{\mathrm{1-loop}} |_{\mathrm{IR}} \sim \mathcal{A}_{\mathrm{tree}} \sum_{i=1}^n 
\frac{(-s_{i,i+1})^{-\epsilon}}{\epsilon^2} \, ,
\end{equation}
which implies that $\mathcal{A}_{\mathrm{tree}}$ is a linear combination of supercoefficients
$\tilde{\mathcal{B}}$ which we just have shown to transform covariantly.
Notice that the only input needed for this alternative proof of tree-level covariance is the knowledge that the three-point MHV and 
$\overline{\rm MHV}$ tree superamplitudes are covariant. Furthermore,  it is not necessary to 
know  the large-$z$ behaviour of the 
superamplitudes.

{\bf 3.} 
Finally, we compare  the remarks of this section to the approach followed by DHKS in \cite{dhks}. 
There, it has been conjectured  that 
a generic  $n$-point amplitude in $\cN=4$ SYM can be written by factoring out the corresponding 
$n$-point MHV amplitude,  as \cite{dhks}
\beq
\cA_n \ = \ \cA_{n, \mathrm{MHV}}  \ \cR
\ , 
\eeq
where $\cR$ is dual superconformal invariant to all loops. 
In the approach outlined here, we have restricted ourselves 
to proving the superconformal covariance of coefficients of the expansion 
of a generic one-loop amplitude in terms of box functions, without separating explicitly 
the (anomalous) MHV superamplitude. It would be interesting to see how this approach 
may provide a link between the  superconformal invariance of the amplitudes as discussed in  \cite{dhks}, 
and  the conformal properties of  the integral functions \cite{magic} appearing in the expansion of 
generic amplitudes in  $\cN=4$ SYM.


\section*{Acknowledgements}

It is a pleasure to thank  Babis Anastasiou, James Drummond, Valeria Gili, 
Valya Khoze, Gregory Korchemsky, Horatiu Nastase, 
Radu Roiban  and especially George Georgiou and Bill Spence for discussions.
We also thank the organisers of the  Florence workshop 
``Strong Coupling: from Lattice to AdS/CFT",  of the  Paris workshop 
``Wonders in Gauge Theory and  Supergravity",  and of the 
Z\"{u}rich ``Workshop on Gauge Theory and String Theory" for the stimulating atmosphere 
of these meetings.  
This work was supported by the STFC under a Rolling Grant PP/D507323/1.
The work of PH is supported by an EPSRC Standard Research Grant EP/C544250/1.
GT is supported by an EPSRC Advanced Research Fellowship EP/C544242/1
and by an EPSRC Standard Research Grant EP/C544250/1.


\end{document}